\newcommand{\NB}{{\it NB816}}
\documentclass{emulateapj}

\shortauthors{Murayama et al.}
\shorttitle{LYMAN$\alpha$ EMITTERS AT $z = 5.7$}

\begin{document}


\title{LYMAN $\alpha$ EMITTERS AT REDSHIFT 5.7 IN THE COSMOS FIELD \footnotemark[1]}

 \author{
	  T. Murayama      \altaffilmark{2},
	  Y. Taniguchi     \altaffilmark{3},
	  N. Z. Scoville   \altaffilmark{4, 5},
	  M. Ajiki         \altaffilmark{2},
	  D. B. Sanders    \altaffilmark{5},
	  B. Mobasher      \altaffilmark{6},
	  H. Aussel        \altaffilmark{5, 7},
         P. Capak         \altaffilmark{4},
	  A. Koekemoer     \altaffilmark{6},
	  Y. Shioya        \altaffilmark{3},
	  T. Nagao         \altaffilmark{8, 9},
         C. Carilli       \altaffilmark{10},
         R. S. Ellis      \altaffilmark{4},
         B. Garilli       \altaffilmark{11},
         M. Giavalisco    \altaffilmark{6},
         M. G. Kitzbichler \altaffilmark{12},
         O. LeFevre       \altaffilmark{13},
         D. Maccagni      \altaffilmark{11},
         E. Schinnerer    \altaffilmark{14},
         V. Smolcic       \altaffilmark{15, 16},
         S. Tribiano      \altaffilmark{17, 18},
         A. Cimatti       \altaffilmark{9},
	  Y. Komiyama      \altaffilmark{8},
	  S. Miyazaki      \altaffilmark{19},
	  S. S. Sasaki     \altaffilmark{2,4},
	  J. Koda          \altaffilmark{4}, and
	  H. Karoji        \altaffilmark{8}
}

\altaffiltext{1}{Based on data collected at the
                 Subaru Telescope, which is operated by
                 the National Astronomical Observatory of Japan}
\altaffiltext{2}{Astronomical Institute, Graduate School of Science,
                 Tohoku University, Aramaki, Aoba, Sendai 980-8578, Japan}
\altaffiltext{3}{Physics Department, Graduate School of Science \&
                 Engineering, Ehime University, 2-5 Bunkyo-cho, Matsuyama
                 790-8577, Japan}
\altaffiltext{4}{Department of Astronomy, MS 105-24, California Institute of
                Technology, Pasadena, CA 91125}
\altaffiltext{5}{Institute for Astronomy,  University of Hawaii,
                 2680 Woodlawn Drive, HI 96822}
\altaffiltext{6}{Space Telescope Science Institute, 3700 San Martin Drive,
                 Baltimore, MD 21218}
\altaffiltext{7}{CEA Saclay, DSM/DAPNIA/SAp, 91191 Gif-sur-Yvette Cedex, France}
\altaffiltext{8}{National Astronomical Observatory of Japan, 2-21-1,
                 Osawa, Mitaka, Tokyo 181-8588, Japan}
\altaffiltext{9}{INAF --- Osservatorio Astrofisico di Arcetri, Largo
                 Enrico Fermi 5, 50125 Firenze, Italy}
\altaffiltext{10}{National Radio Astronomy Observatory,
                  P.O. Box 0, Socorro, NM 87801-0387}
\altaffiltext{11}{INAF, Istituto di Astrofisica Spaziale e Fisica Cosmica,
                  Sezione di Milano, via Bassini 15, 20133 Milano}
\altaffiltext{12}{Max-Planck-Institut f\"ur Astrophysik,
                  D-85748 Garching bei M\"unchen, Germany}
\altaffiltext{13}{Laboratoire d'Astrophysique de Marseille,
                  BP 8, Traverse du Siphon, 13376 Marseille Cedex 12, France}
\altaffiltext{14}{Max Planck Institut f\"ur Astronomie,
                  K\"onigstuhl 17, Heidelberg, D-69117, Germany}
\altaffiltext{15}{Princeton University Observatory, Princeton, NJ 08544}
\altaffiltext{16}{University of Zagreb, Department of Physics, Bijenicka cesta 32, 
                  10000 Zagreb, Croatia}
\altaffiltext{17}{American Museum of Natural History}
\altaffiltext{18}{CUNY Bronx Community College, New York, NY}
\altaffiltext{19}{Subaru Telescope, National Astronomical Observatory of Japan,
                 650 N. A'ohoku Place, Hilo, HI 96720}

\begin{abstract}

We present results from a narrow-band optical survey of a contiguous area 
of 1.95 deg$^2$, covered by the Cosmic Evolution Survey (COSMOS). 
Both optical narrow-band ($\lambda_{\rm c} = 8150$ \AA ~ and 
$\Delta\lambda = 120$ \AA) and broad-band ($B$, $V$, $g^\prime$, $r^\prime$, 
$i^\prime$, and $z^\prime$) imaging observations were performed with the
Subaru prime-focus camera, Suprime-Cam on the Subaru Telescope. 
We provide the largest contiguous narrow-band survey,  
targetting Ly$\alpha$ emitters (LAEs) at $z\approx 5.7$. We find a total of 119
LAE candidates at $z\sim 5.7$. Over the wide-area covered by this survey, we
find no strong evidence for large scale clustering of LAEs. We estimate a
star formation rate (SFR) density of 
$\sim 7 \times 10^{-4} M_\odot$ yr$^{-1}$ Mpc$^{-3}$ for LAEs at $z\approx 5.7$, 
and compare it with previous measurements.
\end{abstract}


\section{INTRODUCTION}

Understanding of the formation and early evolution of galaxies 
requires study of rest-frame properties of well-defined samples of 
high-redshift galaxies. These are needed to address the comsic star formation
history and growth of large scale structures in the early Universe and the 
source of cosmic reionization of intergalactic space. 
There are two widely used techniques to select
such high-redshift galaxies:  (1) Lyman Break technique, aiming for 
Lyman Break Galaxies (LBGs; Steidel et al. 1999;  Iwata et al. 2003; 
Ouchi et al. 2004; Bouwens \& Illingworth 2006 and references therein), and 
(2) narrow-band imaging surveys, targeting Ly$\alpha$ emitters (LAEs; 
Hu \& McMahon 1996; Rhoads \& Malhotra 2001; Ajiki et al. 2003;
Hu et al. 2004; Taniguchi et al. 2005 and references therein). The narrow-band
surveys are mainly aimed at star-forming population while Lyman Break technique
also selects galaxies with older age.

Recently, LAE surveys have extended study of the clustering and morphology of
star-forming galaxies to the highest redshifts (Shimasaku et al. 2004;
Ouchi et al. 2005; Ajiki et al. 2006; Mobasher et al. 2006).
Indeed, some of the previous LAE surveys have shown signs of large scale
structures, either using 2-D projected distribution of galaxies or 3-D 
distribution, also using redshifts. For example, evidence for clustering of
LAEs at $z \approx 4.9$ was found over an area of 
$\simeq$ 0.5 degree $\times$ 0.5 degree, covered by the Subaru Deep Field 
(SDF) (Shimasaku et al. 2004; see also Ouchi et al. 2003), extending to
$\sim$ 20 Mpc $\times$ 50 Mpc. However, no such structures were found for
LAEs at $z \approx 5.7$
(Shimasaku et al. 2006) and $z \approx 6.6$ (Taniguchi et al. 2005;
Kashikawa et al. 2006). Although, using a spectroscopically confirmed sample
of 34 LAEs at $z \approx 5.7$, Shimasaku et al. (2006) found evidence for
weak clustering. In an independent study, using a spectroscopically confirmed 
sample of 19 LAEs at $z \approx 5.7$, Hu et al. (2004) found structures
extending to angular scales of $\sim 60$ Mpc with evidence for 
filamentary structures. This result was further confirmed by 
Ouchi et al. (2005) 
who found filamentary structures of size 10 -- 40 Mpc
in the Subaru XMM-Newton Deep Survey (SXDS: Sekiguchi et al. 2004), using
a photometric sample of 515 LAEs. 
In addition to the above results, based on deep narrow-band surveys
in so-called ``blank fields'', evidence has been accumulating in support
of clustering at $z = 4.1$ to 5.2, with a few Mpc scales around 
high-redshift radio galaxies (Venemans et al. 2002, 2004; Overzier
et al. 2006) and quasar SDSS J0836+0054  
(Zheng et al. 2006; Ajiki et al. 2006b). 

These studies are useful in investigating early formation of galaxies 
and large-scale
structures. Furthermore, they provide important constraints on both
the star formation and cosmic re-ionization history. A detailed study of
clustering of galaxies at high-redshifts requires deep and
wide-area surveys to allow a homogeneously selected sample of galaxies
and to minimise effects of cosmic variance. This is the subject of the
present paper.

In this paper we present the largest survey of
LAEs at $z \approx 5.7$, covering the entire 2 square degree field of
the Cosmic Evolution Survey (COSMOS), centered at 
$\alpha$(J2000.0) = $10^{\rm h} ~ 00^{\rm m} ~ 28.6^{\rm s}$ and
$\delta$(J2000.0) = $+02^\circ ~ 12' ~ 21.0''$ (Scoville et al. 2007).
The full COSMOS field has been observed in $I_{814}$-band with
the Advanced Camera for Surveys (ACS) on-board the Hubble Space 
Telescope (HST). In addition to the ACS data, the multi-wavelength broad 
and narrow-band observations were also performed using the
Supreme-Cam (Miyazaki et al. 2002) on the Subaru Telescope (Kaifu et al. 2000; Iye et al. 2004). The narrow-band filter \NB~ has an
effective wavelength of 
$\lambda_c = 8150$ \AA~ with a width $\Delta\lambda =120$ \AA~ 
(see Ajiki et al. 2003 for details), allowing to select LAEs in the range 
$ 5.65 < z < 5.75$. We assume standard cosmology with 
$\Omega_{\rm matter} = 0.3$, $\Omega_\Lambda = 0.7$,
and $H_0 = 70$ km s$^{-1}$ Mpc$^{-1}$.
Throughout this paper, we use magnitudes in the AB system.

\section{OBSERVATIONS AND SAMPLE SELECTION}

\subsection{Data and Source Detection}

We carried out an optical narrow-band (\NB) imaging survey of the entire
2-deg$^2$ area of the COSMOS field, using the Suprime-Cam on the Subaru 
Telescope.  These observations, combined with the broad-band 
($B$, $V$, $g^\prime$, $r^\prime$, $i^\prime$,
and $z^\prime$) Suprime-Cam and ACS ($I_{814}$) photometric data 
will be used to identify LAE candidates at $z\sim 5.7$ and to study their 
properties. Details of the narrow-band and ground-based   
observations and data reduction are given in Taniguchi et al. (2007)
and Capak et al. (2007a) and for the HST-ACS observations in
Koekemoer et al. (2007). 

For the ground-based observations, 
the seeing size varies between the exposures.
The PSF size of each \NB{} images is
between $0\farcs{}4$ and $0\farcs{}7$.
To optimize source detection in \NB{}, 
we only use exposures with PSF sizes smaller than $1\farcs{}15$
to construct a combined image. 

The limiting magnitude of the \NB{} has a variance of $\sim 0.5$ mag,
depending on
the location on the image. We only confine our LAE search to areas
with low noise where the 5$\sigma$ limiting magnitude
with a $2''$ diameter aperture in \NB{} is $\sim 25.1$.  
This corresponds to a total effective area of 1.95 deg$^2$.
The transverse co-moving area of the LAE survey at $z=5.7$ is
$3.9\times 10^4$ Mpc$^2$. The FWHM of the filter, which has a
Gaussian-like shape, corresponds to a co-moving depth of 45 Mpc, 
spanning the redshift range $5.65 < z < 5.75$ along the line-of-sight.
Therefore, our \NB~ survey probes a volume of $1.8\times 10^6$ Mpc$^3$.

We have performed source detection and photometry on the \NB{} image 
with SExtractor version
2.3 (Bertin \& Arnouts 1996). A source is selected as a contiguous 9-pixel 
area above the 3 $\sigma$ noise level (corresponding to 26.48 
mag arcsec$^{-2}$) on the \NB{} image.
Photometry is  performed on the \NB~ and 
the broadband images over $0\farcs{}5$, $2''$, and $3''$ diameter apertures.
To the magnitude limit of the survey
[$\NB(2''\phi) = 25.1$], we find $\sim 3\times 10^5$ sources.

\subsection{Selection of LAE Candidates}

\begin{figure}
\plotone{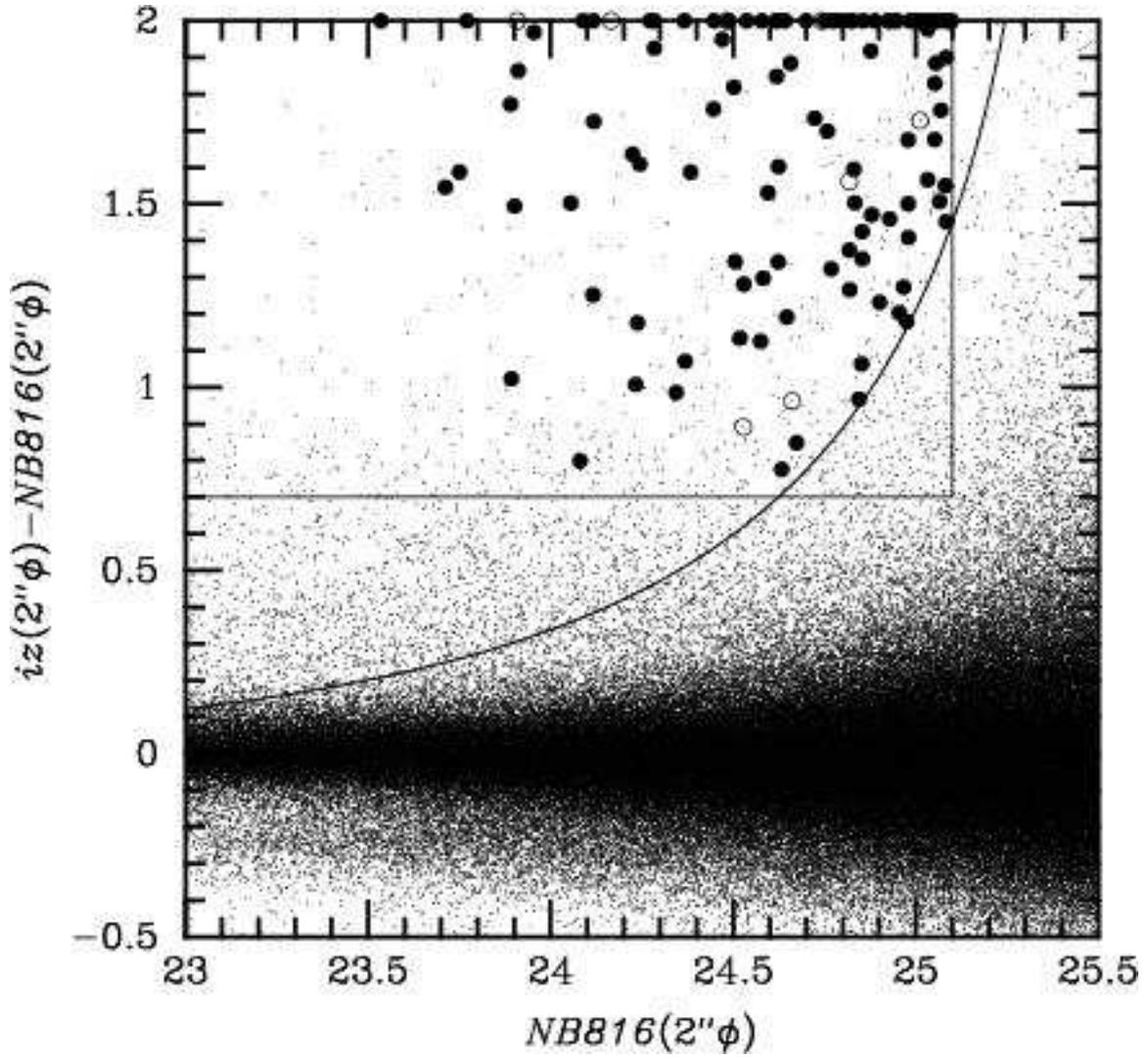}
\caption{Diagram between $iz-\NB$ and \NB ~  for all objects detected with $\NB < 26$.
         Our LAE candidates are shown by filled circles (the statistical sample)
         and open circles (the non-statistical sample).
         Objects with $iz-\NB > 2$ are shown at $iz-\NB = 2$ for clarity.
         The vertical line shows the detection limit of \NB$=25.1$.
         The horizontal line shows the selection criterion of $iz- \NB= 0.7$.
         The curve shows the 3 $\sigma$ limit for the $iz-\NB$ color.
\label{nbcm}}
\end{figure}

Our main aim in the present survey is to identify reliable LAE candidates
at $z \approx 5.7$ by first selecting \NB{} excess objects, 
using the following criteria:

\begin{eqnarray}
      \NB (2''\phi) & < & 25.1, \\
 iz (2''\phi) - \NB (2''\phi) & > & max( 0.7, 3\sigma_{iz-{\NB}}), \label{cr:ex}\\
         B(0\farcs 5\phi)  & > & 29.6,\label{cr:B}\\
  g^\prime(0\farcs 5\phi)  & > & 29.2,\label{cr:g}\\
         V(0\farcs 5\phi)  & > & 29.1,\label{cr:V} \;\;{\rm and}\\
  r^\prime(0\farcs 5\phi)  & > & 29.1,\label{cr:r}
\end{eqnarray}
where $iz$ is the continuum magnitude at $\lambda = 8150$ \AA, estimated 
by linear interpolation between $i^\prime$ and $z^\prime$ flux densities 
($f_{iz}= 0.57 f_{i^\prime} + 0.43 f_{z^\prime}$).
The first criterion ensures that objects are detected above the 5 $\sigma$ 
level in the \NB{}.
The criterion \ref{cr:ex} allows selection of emission-line objects
with observed equivalent width, $EW_{\rm obs} \geq 120$ \AA.
The $3\sigma$ of $iz - \NB$ is estimated from
the local-background noise measurement of the most noisy region in the survey 
area.
This is illustrated in Figure \ref{nbcm} where the LAE candidates are
identified on $iz -  \NB$ {\it vs.} \NB~ color---magnitude diagram.
The criteria (\ref{cr:B}), (\ref{cr:g}), (\ref{cr:V}), and (\ref{cr:r}) 
ensure that $z\approx 5.7$ candidates are undetected
(at $\approx 1.5 \sigma$ noise level) in $B$, $g^\prime$, $V$, and $r^\prime$
bands.  We adopt magnitudes in a 0\farcs{}5 diameter aperture
to avoid possible contamination by low-$z$ foreground objects.

We find a total of 119 LAE candidates at $z\sim 5.7$ that satisfy the 
above criteria and confirm these by careful eye inspection for apparent false 
detections. Our first spectroscopic followup observation 
of $\sim$50 LAE candidates indicates a $\sim$95\% or better confirmation rate
(Capak et al. 2007b).  
We also find that there is no low-luminosity radio-loud AGN with
$L({\rm 1.4GHz})>6\times10^{24}$ W Hz$^{-1}$ among our final LAE
candidates (Carilli et al. 2007). 
The coordinates and photometric data for the LAE candidates are listed
in Table \ref{tab:LAE}.  
All magnitudes are corrected for Galactic extinction; 
$\bar{E}(B - V) = 0.0195$ (Capak et al. 2007a). 
In Table \ref{tab:LAE} we also list emission line fluxes and
observed equivalent widths estimated
from \NB{} and $z^{\prime}$ flux densities in a $3''$ aperture.

As shown by the criterion \ref{cr:ex},
the detection limit of LAEs depends on the depth of both the \NB{} 
and $i^{\prime}$ and $z^{\prime}$ band images. 
However, we did not impose any noise threshold on the $i^{\prime}$ 
and $z^{\prime}$ band images 
when we selected the LAE candidates.  
To ensure homogeneity, we make a subsample of the LAE candidates  
in areas with low noise, both in the $i^{\prime}$ and $z^{\prime}$ bands 
where the 5$\sigma$ limiting magnitude over a $2''$ diameter aperture
is brighter than 25.4 or 24.8 mag in $i^{\prime}$ and $z^{\prime}$ bands
respectively.  
This subsample (hereafter, the statistical sample) contains
111 LAE candidates which are identified by asterisks in Table \ref{tab:LAE}. 
We use this ``statistical sample'' to estimate statistical properties of LAEs, 
including their number density and luminosity function. 
The size of the effective area corresponding to this statistical sample
is reduced to 1.86 deg$^2$, equivalent to $3.7\times 10^4$ Mpc$^2$ of
the transverse co-moving area at $z$=5.7. 
The survey volume for the statistical sample is $1.7\times 10^6$ Mpc$^3$.

\section{Results and Discussion}

\subsection{Spatial Distribution and Angular Two-Point Correlation Function}

In Figure \ref{map}, we show the spatial distribution of all the LAE 
candidates in our sample.  
There appears to be little evidence for strong clustering.
When dividing the survey area into four tiles with
0.7 degree $\times$ 0.7 degree each, we find 26, 28, 26, and 31 LAEs
in the SW, NW, SE, and NE quadrant respectively. Therefore, 
the 111 LAE candidates in the statistical sample are almost randomly
distributed at least in large scale.

We derive the angular two-point correlation function (ACF), $\omega(\theta)$,
for the 111 LAE candidates in the ``statistical sample''
using the estimator defined by Landy \& Szalay (1993)
\begin{equation}
 w(\theta) = \frac{DD(\theta)-2DR(\theta)+RR(\theta)}{RR(\theta)},
 \label{two-point}
\end{equation}
where $DD(\theta)$, $DR(\theta)$, and $RR(\theta)$ are normalized numbers of
galaxy-galaxy, galaxy-random, and random-random pairs, respectively.
The random sample consists of 100,000 sources with the same geometrical
constraints as the LAE sample.

In the left panel of Figure \ref{acf},
we show the ACF for the statistical LAE sample (large filled circles).
There is a clustering signal with  2.8$\sigma$ significance
at 25 arcsec ($\simeq$ 1.0 Mpc at $z=5.7$).
We also find possible signals at 63 arcsec, 100 arcsec, 158 arcsec, and 251 arcsec
although their detection significance is as low as $\simeq$1.3$\sigma$. 
We compare this with the ACF for LAE candidates at $z\approx 5.7$ in the SDF 
(Shimasaku et al. 2006) using their whole sample; $\NB \le 26.5$ 
(small filled circles) and their bright sample; $\NB \le 25.5$ (open circles). 
Most of data points are consistent within 1$\sigma$ errors among the 
three samples. 
However, there seems to be possible difference among the three samples 
at small angular scales.  
Since our LAE sample is limited at $\NB < 25.1$, the brighter sample seems 
to show the larger $w(\theta)$ at $\theta \le 30$ arcsec.

Although the clustering signals are not high,
particularly at larger scales (e.g., $\theta\ge 100$ arcsec),
the LAE candidates in the COSMOS field follow a power-law relation
between $w(\theta)$ and $\theta$, as shown in the right panel 
(i.e.  $w(\theta) = A_w \theta^\beta$). 
We find a best fit power law with an index of 
$\beta  = -1.2 \pm 0.2$ and an amplitude
at 1 arcsec of $A_w = 95\pm 76$.
For this power law fit, we only used data points between 
25 arcsec  and 1585 arcsec except the point at 398 arcsec.
The value of $\beta$ is steeper than $\beta  = -0.74$ (at $z = 4$) 
and $\beta  = -0.81$ (at $z=5$) found for Lyman break galaxies in the SDF
(Kashikawa et al. 2006).  
The semi-analytic models predict that 
the ACF for higher halo mass has a steeper slope and a stronger amplitude 
(Kashikawa et al. 2006), our results imply that the LAEs are embedded in 
massive dark matter halo, 
e.g., heavier than $10^{12}M_\odot$ (Ouchi et al. 2004). However, absence of 
significant clustering signal 
for LAEs at $z \approx 5.7$ in the SDF (Shimasaku et al. (2006))
may be indicative of cosmic variance in these studies. 

\subsection{Ly$\alpha$ Luminosities of LAE Candidates}

Using our total sample of LAE candidates at $z\approx5.7$ 
and a volume of $1.7 \times 10^6$ Mpc$^3$, we estimate a space density of 
 $n$(Ly$\alpha$)$\approx6.6 \times 10^{-5}$ Mpc$^{-3}$ for the LAE candidates 
in our ``statistical sample''. This is a factor of 1.6 lower than those
in previous LAE surveys with nearly the same \NB~ limits (\NB~$\approx 25$)
(Ajiki et al. 2003; Hu et al. 2004). The observed difference is likely caused
by cosmic variance.

We measure the Ly$\alpha$ luminosity, $L$(Ly$\alpha$),
of our LAE candidates, assuming all the LAEs to be at $z=5.70$. 
The rest-frame equivalent widths and Ly$\alpha$ luminosities as listed in  
Table \ref{tab:LLAE}. We adopted aperture magnitudes measured over 
a $3''$ aperture diameter for estimation of the Ly$\alpha$ luminosity. 
We find that the magnitudes measured over $2''$ and $3''$ 
apertures are well correlated with a small systematic offset.
However, the total magnitudes from SExtractor (MAGAUTO) 
show large scatter when compared with the aperture magnitudes measured over 
$2''$ and $3''$ diameter apertures due to contamination by neighboring objects.
The derived Ly$\alpha$ luminosities range from $\approx 6.3 \times 10^{42} $ 
ergs s$^{-1}$ to $\approx 3.1 \times 10^{43} $ ergs s$^{-1}$.  

Using the 111 LAE candidates in our ``statistical sample'', we constructed the 
Ly$\alpha$ luminosity 
function, as shown in the left panel of Figure \ref{ll}. We compare this
with other Ly$\alpha$ luminosity functions at $z\sim 5.7$
(Rhoads \& Malhotra 2001; Ajiki et al. 2003, 2006a; Hu et al. 2004).  
Our survey has a luminosity limit of 
$L_{\rm lim}$(Ly$\alpha$)$=6.3 \times 10^{42} $ ergs s$^{-1}$, 
corresponding to a LAE with $\NB=25.1$, undetected in $i'$ and in $z'$.  
This limit is similar to those in Ajiki et al. (2003) and 
Hu et al. (2004). 
The luminosity functions do not show
any significant difference at the luminous part 
[$L$(Ly$\alpha$)$ \ge 10^{43} $ ergs s$^{-1}$],
although the number density of Rhoads \& Malhotra (2001) is smaller 
compared to those in other surveys. Because of the improved Poisson statistics
in the COSMOS LAE sample, the errorbars in this survey are relatively smaller. 
Therefore, the brighter part of the LAE luminosity function is relatively well
established. Also, the wide area covered by the COSMOS survey enables us 
to improve the statistics at the brighter part of the luminosity function
($L$(Ly$\alpha$) $> 2 \times 10^{43} $ ergs s$^{-1}$)
by increasing the number of luminous LAEs per luminosity bin compared 
to previous surveys.
At $10^{42.8} \le L$(Ly$\alpha$)$ < 10^{43.0}$ ergs s$^{-1}$, 
the number density from this study is almost the same as that of
Rhoads \& Malhotra (2001) and Ajiki et al. (2006a)
while is smaller (by a factor of 2) than that estimated in Hu et al. (2004) 
and Ajiki et al. (2003). This is, at least partly, due to both
spatial variance and incompleteness 
at $10^{42.8} \le L$(Ly$\alpha$)$ < 10^{43.0}$ ergs s$^{-1}$ in our survey.

In Figure \ref{ll} (right panel), 
we also compare our result with those of previous LAE surveys
as a function of redshift. The LAE luminosity functions  
at $z=3.4$, $z=4.9$, and $z=6.6$ are taken from Cowie \& Hu (1998), 
Ouchi et al. (2003), and Taniguchi et al. (2005), respectively. 
COSMOS ($z=5.7$) and Taniguchi et al. 2005 ($z=6.6$) give
almost the same result within errors at the bright end 
($L$(Ly$\alpha$)$ \ge 10^{43}$ ergs s$^{-1}$). 
No LAE with $L$(Ly$\alpha$)$ \ge 10^{43}$ ergs s$^{-1}$ 
is found in the previous LAE surveys at $z=3.4$ and $z=4.9$.
While there are differences on the detection completeness and the contamination
rates among the surveys,
this may provide evidence for the number and/or luminosity
evolutions of LAEs.

\subsection{Star Formation Rates}

We now estimate the contribution from LAEs to the star formation rate (SFR) 
at $z \approx 5.7$, using the following relation 
(Kennicutt 1998; Brocklehurst 1971),
\begin{equation}
SFR({\rm Ly}\alpha)  = 9.1 \times 10^{-43} L({\rm Ly}\alpha) ~ M_\odot ~ {\rm yr}^{-1},
\end{equation}
where $L($Ly$\alpha)$ is in ergs s$^{-1}$. We assume 
Salpeter initial mass function with ($m_{\rm lower}$, $m_{\rm upper}$)
= (0.1 $M_\odot$, 100 $M_\odot$). 

The estimated SFRs are given in the forth column of Table \ref{tab:LLAE}.
They range from 5.7 to 28.3 $ M_\odot$ yr$^{-1}$
with a median value of 9.6 $M_\odot$ yr$^{-1}$.  
The SFRs derived here can be underestimated due to the effect of
absorption by H {\sc i} gas both in the host galaxies and in 
the intergalactic medium and dusts in the host galaxies. 
SFRs independent of the absorption are estimated by the radio data;
an upper limit to the mean massive star
formation rate (5 $M_\odot$ to 100 $M_\odot$) for the LAE sample
is derived as $\sim 100$ $M_\odot$ yr$^{-1}$ (Carilli et al. 2007).

The estimated SFRs from the Ly$\alpha$ luniomsities here are comparable to those of LAEs at 
$z \simeq$ 5.7 -- 6.6 (e.g., Ajiki et al. 2003; Taniguchi et al. 2005).
The SFR density at $z=5.7$ is estimated by summing up the Ly$\alpha$ 
luminosities of the 111 ``statistical candidates'' and corresponds to 
$7.2\times 10^{-4}$ $M_{\odot}$ yr$^{-1}$ Mpc$^{-3}$, 
similar to those obtained in previous narrow-band surveys.

It would be instructive to examine whether the SFRs derived from the 
Ly$\alpha$ luminosity here are consistent with those based on
the UV continuum luminosity from the broad-band data. 
The observed $z^\prime$ magnitudes, measured over $3''$ aperture diameters, 
are converted to UV continuum luminosities at $\lambda = 1270$ \AA and used to
estimate the SFRs, using the relation 
(Kennicutt 1998; see also Madau et al. 1998),
\begin{equation}
\label{UVtoSFR}
SFR({\rm UV}) = 1.4 \times 10^{-28}L_{\nu}
~~ M_{\odot} ~ {\rm yr}^{-1},
\end{equation}
where $L_\nu$ is the UV continuum luminosity in 
ergs s$^{-1}$ Hz$^{-1}$. For each object, 
we estimate the SFR from its rest-frame UV ($\lambda=1270$\AA)
continuum luminosity, with the results summarized
in Table \ref{tab:LLAE}. 
Comparison between $SFR$(Ly$\alpha$) and $SFR$(UV) in Figure \ref{sfr}   
shows that, on average, $SFR$(UV) is relatively higher than $SFR$(Ly$\alpha$) 
for most of the LAE candidates. We find an average ratio of
$SFR_{\rm total}$(Ly$\alpha$) / $SFR_{\rm total}$(UV)$ = 0.68$,
where $SFR_{\rm total}$(Ly$\alpha$) and $SFR_{\rm total}$(UV) are, 
respectively, the sum of SFRs of all our 119 LAE candidates 
from Ly$\alpha$ line and $UV$ continuum. 
The relatively lower $SFR$(Ly$\alpha$) compared to $SFR$(UV)
is likely due to the effect of the differential absorption.
However, some of our LAE candidates have
$SFR$(Ly$\alpha$)/$SFR$(UV) $>1$, e.g., those ratios of
\#13, \#27, \#50, \#71, and \#114
are greater than unity with 2$\sigma$ significance.
They may be in a very early phase ($<10^8$ yr) of star formation activity
in which  $SFR$(UV) is underestimated (Schaerer 2000; 
see also Nagao et al. 2004, 2005).

\section{SUMMARY}

We have presented results from our narrow-band deep imaging survey
of the COSMOS field, targetting LAEs at $z \approx 5.7$. 
This is the largest contiguous survey of LAEs. Our main results are 
summarized below:

(1) We found 119 LAE candidates in our narrow-band survey. Allowing for
changes in the
noise level over the entire field, we extracted a subsample of 111 LAEs
that is used to make our statistical analysis.

(2) We find no significant evidence for clustering of LAEs at $z\sim 5.7$
contrary to some of the previous LAE surveys.
 
(3) An analysis of angular two-point correlation function gives the power-law 
relation $w(\theta) = A_w \theta^\beta$, with $\beta  = -1.2 \pm 0.2$. 
The power-law index here is steeper than that found for Lyman break galaxies 
at $z =4$ and 5. 
This suggests that LAEs at $z \approx$ 5.7 might be
located in massive dark matter halos with mass of $> 10^{12} M_\odot$.

(4) The number density of LAEs and the average star formation rate are similar
to those measured in previous surveys. We estimate a star formation rate 
density of
$\sim 7 \times 10^{-4} M_\odot$ yr$^{-1}$ Mpc$^{-3}$ at $z \approx 5.7$.

(5) We measure the Ly$\alpha$ luminosity function at $z >5$ and 
extend this to  
$L$(Ly$\alpha$)$ \ge 10^{43}$ ergs s$^{-1}$, not explored by 
previous LAE surveys. We compare the estimated  Ly$\alpha$ luminosity function
here with those in previous studies in the range $z = 3.4- 6.6$.

\acknowledgements

We would like to thank both the Subaru and HST staff for their invaluable help.
We also thank Masami Ouchi for providing us his data.
This work was financially supported in part by the Ministry
of Education, Culture, Sports, Science, and Technology (Nos. 10044052 
and 10304013), and by JSPS (15340059 and 17253001). SSS and TN
are JSPS fellows.

%

\clearpage

\LongTables
\begin{deluxetable}{rcccccccccccc}
\tabletypesize{\scriptsize}
\tablenum{1}
\tablecaption{Photometric properties of the LAE candidates at $z\approx 5.7$\label{tab:LAE}}
\tablewidth{0pt}
\tablehead{
 \colhead{No.\tablenotemark{a}} &
 \colhead{$\alpha$(J2000)} &
 \colhead{$\delta$(J2000)} &
 \colhead{{\it NB816}\tablenotemark{b}} &
 \colhead{$i^\prime$\tablenotemark{b}} &
 \colhead{$z^\prime$\tablenotemark{b}} &
 \colhead{$iz$\tablenotemark{b}} &
 \colhead{{\it NB816}\tablenotemark{b}} &
 \colhead{$z^\prime$\tablenotemark{b}} &
 \colhead{$f_\nu ({\it NB816})$\tablenotemark{c}} &
 \colhead{$f_\nu (z^\prime)$\tablenotemark{c}} &
 \colhead{$f_{\rm line}$\tablenotemark{d}} &
 \colhead{$EW_{\rm obs}$\tablenotemark{e}} \\
 &
 ($\arcdeg$)  &
 ($\arcdeg$) &
 ($\phi$2\farcs 0) &
 ($\phi$2\farcs 0) &
 ($\phi$2\farcs 0) &
 ($\phi$2\farcs 0) &
 ($\phi$3\farcs 0) &
 ($\phi$3\farcs 0) &
 ($\phi$3\farcs 0) &
 ($\phi$3\farcs 0) &
 &
 (\AA)
}

\startdata
1* & 149.4786 & +2.2116 & 24.9 & 26.3 & 26.4 & 26.3 & 24.5 & 25.9 & $5.6 \pm 0.6$ & $1.6 \pm 0.9$ & $2.6 \pm  0.4$ & $353 \pm 207$ \\
2* & 149.5460 & +2.7640 & 24.8 & 26.7 & 26.0 & 26.3 & 24.6 & 25.9 & $5.0 \pm 0.7$ & $1.6 \pm 1.0$ & $2.3 \pm 0.5$ & $307 \pm 191$ \\
3* & 149.5790 & +1.5183 & 25.1 & 26.9 & 26.2 & 26.5 & 24.7 & 26.3 & $4.8 \pm 0.7$ & $1.1 \pm 1.0$ & $2.3 \pm 0.5$ & $452 \pm 401$ \\
4* & 149.6052 & +2.4476 & 23.7 & 25.2 & 25.3 & 25.2 & 23.2 & 24.8 & $18.3 \pm 0.6$ & $4.3 \pm 1.0$ & $8.8 \pm 0.4$ & $452 \pm 107$ \\
5* & 149.6320 & +2.3466 & 24.7 & 27.1 & 26.7 & 26.9 & 24.4 & 26.2 & $6.2 \pm 0.8$ & $1.2 \pm 1.0$ & $3.0 \pm 0.5$ & $578 \pm 498$ \\
6* & 149.6331 & +1.8223 & 25.0 & 27.7 & 27.3 & 27.5 & 24.7 & 26.4 & $4.7 \pm 0.7$ & $<1.0$       & $2.3 \pm 0.5$ & $>498$ \\
7* & 149.6361 & +2.3276 & 24.9 & 26.5 & 26.3 & 26.4 & 24.6 & 25.6 & $5.2 \pm 0.7$ & $2.2 \pm 0.9$ & $2.2 \pm 0.4$ & $227 \pm 105$ \\
8* & 149.6537 & +1.5394 & 24.7 & 26.1 & 25.0 & 25.5 & 24.2 & 24.8 & $7.4 \pm 0.7$ & $4.5 \pm 1.0$ & $2.8 \pm 0.5$ & $136 \pm 39$ \\
9* & 149.6804 & +2.7076 & 24.8 & 27.2 & 27.2 & 27.2 & 24.6 & 30.4 & $5.4 \pm 0.8$ & $<1.0$ & $2.9 \pm 0.5$ & $>659$ \\
10* & 149.6886 & +2.9180 & 24.5 & 26.0 & 25.3 & 25.6 & 24.1 & 24.6 & $8.7 \pm 0.7$ & $5.0 \pm 1.1$ & $3.3 \pm 0.5$ & $147 \pm 39$ \\
11 & 149.7334 & +1.4765 & 24.7 & 25.7 & 25.5 & 25.6 & 24.1 & 24.8 & $8.6 \pm 0.9$ & $4.5 \pm 1.4$ & $3.4 \pm 0.6$ & $171 \pm 61$ \\
12* & 149.7362 & +1.5594 & 25.0 & 27.9 & 99.0 & 99.0 & 24.7 & 99.0 & $4.6 \pm 0.8$ & $<1.2$ & $2.5 \pm 0.5$ & $>463$ \\
13* & 149.7457 & +2.7519 & 24.1 & 27.1 & 27.4 & 27.2 & 23.9 & 99.0 & $9.6 \pm 0.7$ & $<1.0$ & $5.2 \pm 0.5$ & $>1172$ \\
14* & 149.7493 & +2.7609 & 24.1 & 25.6 & 25.4 & 25.5 & 23.6 & 25.0 & $12.9 \pm 0.7$ & $3.6 \pm 1.0$ & $6.0 \pm 0.5$ & $369 \pm 105$ \\
15* & 149.7537 & +2.2380 & 24.8 & 26.3 & 25.4 & 25.8 & 24.6 & 25.1 & $5.1 \pm 0.7$ & $3.3 \pm 0.9$ & $1.9 \pm 0.5$ & $127 \pm 47$ \\
16* & 149.7725 & +1.7967 & 25.1 & 27.2 & 27.7 & 27.4 & 24.9 & 27.5 & $3.9 \pm 0.7$ & $<1.0$ & $2.0 \pm 0.5$ & $>454$ \\
17* & 149.7810 & +1.5176 & 24.6 & 25.9 & 25.8 & 25.9 & 24.2 & 25.2 & $7.6 \pm 0.9$ & $3.1 \pm 1.1$ & $3.3 \pm 0.6$ & $237 \pm 96$ \\
18* & 149.7812 & +1.4940 & 24.1 & 25.8 & 25.8 & 25.8 & 23.8 & 25.7 & $10.9 \pm 0.9$ & $2.0 \pm 1.2$ & $5.4 \pm 0.6$ & $612 \pm 385$ \\
19* & 149.8019 & +1.8274 & 24.6 & 25.7 & 26.3 & 25.9 & 24.2 & 25.9 & $7.6 \pm 0.7$ & $1.6 \pm 1.1$ & $3.7 \pm 0.5$ & $517 \pm 360$ \\
20* & 149.8023 & +2.2251 & 25.0 & 27.0 & 26.2 & 26.6 & 24.7 & 26.0 & $4.9 \pm 0.6$ & $1.5 \pm 1.0$ & $2.2 \pm 0.4$ & $344 \pm 241$ \\
21* & 149.8082 & +2.6360 & 25.1 & 27.5 & 27.0 & 27.3 & 24.8 & 27.5 & $4.6 \pm 0.8$ & $<1.0$ & $2.4 \pm 0.5$ & $>529$ \\
22 & 149.8085 & +1.5439 & 24.8 & 26.2 & 26.5 & 26.4 & 24.3 & 99.0 & $6.7 \pm 0.8$ & $<1.6$ & $3.6 \pm 0.6$ & $>498$ \\
23* & 149.8186 & +1.7204 & 25.0 & 26.1 & 26.2 & 26.1 & 24.4 & 25.9 & $6.1 \pm 0.7$ & $1.6 \pm 1.0$ & $2.8 \pm 0.5$ & $387 \pm 237$ \\
24* & 149.8203 & +2.7823 & 24.5 & 26.6 & 26.4 & 26.5 & 24.3 & 26.4 & $6.9 \pm 0.7$ & $<1.1$ & $3.5 \pm 0.5$ & $>669$ \\
25* & 149.8323 & +2.0561 & 25.1 & 26.9 & 26.5 & 26.7 & 24.6 & 26.7 & $5.3 \pm 0.6$ & $<0.9$ & $2.6 \pm 0.4$ & $>633$ \\
26* & 149.8380 & +1.6948 & 25.1 & 26.8 & 30.2 & 27.4 & 25.1 & 28.7 & $3.5 \pm 0.6$ & $<0.9$ & $1.8 \pm 0.4$ & $>473$ \\
27* & 149.8443 & +2.7615 & 24.3 & 26.2 & 26.2 & 26.2 & 24.0 & 25.8 & $9.4 \pm 0.5$ & $1.8 \pm 0.7$ & $4.6 \pm 0.4$ & $571 \pm 239$ \\
28* & 149.8466 & +2.7517 & 25.1 & 26.8 & 99.0 & 99.0 & 24.8 & 27.0 & $4.3 \pm 0.6$ & $<0.9$ & $2.2 \pm 0.4$ & $>518$ \\
29* & 149.8776 & +2.3317 & 24.9 & 25.9 & 26.4 & 26.1 & 24.4 & 26.0 & $6.5 \pm 0.6$ & $1.4 \pm 1.0$ & $3.1 \pm 0.4$ & $481 \pm 328$ \\
30* & 149.8893 & +2.8322 & 24.6 & 26.4 & 26.8 & 26.6 & 24.3 & 25.8 & $7.2 \pm 0.7$ & $1.7 \pm 1.1$ & $3.5 \pm 0.5$ & $445 \pm 286$ \\
31* & 149.9107 & +1.6147 & 24.1 & 26.3 & 26.0 & 26.2 & 23.8 & 26.2 & $11.2 \pm 0.7$ & $1.2 \pm 0.9$ & $5.7 \pm 0.4$ & $1031 \pm 767$ \\
32 & 149.9197 & +1.4827 & 24.2 & 26.9 & 25.8 & 26.3 & 23.9 & 26.1 & $10.3 \pm 0.9$ & $1.4 \pm 1.2$ & $5.2 \pm 0.6$ & $851 \pm 748$ \\
33* & 149.9303 & +1.5980 & 24.5 & 26.3 & 25.4 & 25.8 & 24.2 & 25.1 & $7.5 \pm 0.8$ & $3.5 \pm 1.0$ & $3.1 \pm 0.5$ & $199 \pm 66$ \\
34* & 149.9336 & +2.0141 & 25.0 & 26.9 & 26.0 & 26.4 & 24.7 & 25.5 & $5.0 \pm 0.6$ & $2.4 \pm 1.0$ & $2.1 \pm 0.4$ & $191 \pm 88$ \\
35* & 149.9422 & +2.1286 & 24.8 & 26.1 & 26.0 & 26.1 & 24.4 & 25.6 & $6.1 \pm 0.6$ & $2.1 \pm 0.8$ & $2.7 \pm 0.4$ & $283 \pm 120$ \\
36* & 149.9447 & +1.5357 & 24.8 & 27.2 & 25.9 & 26.4 & 24.5 & 25.9 & $5.7 \pm 0.7$ & $1.6 \pm 1.0$ & $2.6 \pm 0.5$ & $375 \pm 241$ \\
37* & 149.9586 & +2.9017 & 24.6 & 27.0 & 27.4 & 27.2 & 24.4 & 99.0 & $6.4 \pm 0.7$ & $<1.2$ & $3.5 \pm 0.5$ & $>667$ \\
38* & 149.9625 & +2.5397 & 24.8 & 26.9 & 30.2 & 27.5 & 24.5 & 99.0 & $5.6 \pm 0.7$ & $<1.0$ & $3.1 \pm 0.5$ & $>670$ \\
39* & 149.9672 & +1.6231 & 24.4 & 26.2 & 26.1 & 26.2 & 24.2 & 26.0 & $7.8 \pm 0.6$ & $1.4 \pm 0.9$ & $3.8 \pm 0.4$ & $596 \pm 376$ \\
40* & 149.9719 & +2.1182 & 24.2 & 25.9 & 24.7 & 25.2 & 23.9 & 24.4 & $10.0 \pm 0.6$ & $6.1 \pm 1.0$ & $3.8 \pm 0.4$ & $137 \pm 27$ \\
41* & 149.9735 & +2.8166 & 24.6 & 27.2 & 28.3 & 27.5 & 24.5 & 99.0 & $5.8 \pm 0.7$ & $<1.0$ & $3.2 \pm 0.5$ & $>690$ \\
42* & 149.9772 & +2.2546 & 24.9 & 27.1 & 27.1 & 27.1 & 24.8 & 27.3 & $4.4 \pm 0.6$ & $<1.1$ & $2.3 \pm 0.4$ & $>479$ \\
43* & 149.9783 & +2.1776 & 24.5 & 26.5 & 26.3 & 26.4 & 24.2 & 26.1 & $7.5 \pm 0.6$ & $1.4 \pm 1.0$ & $3.7 \pm 0.4$ & $608 \pm 436$ \\
44* & 149.9792 & +1.7890 & 24.6 & 26.6 & 26.3 & 26.5 & 24.3 & 26.0 & $6.8 \pm 0.6$ & $1.4 \pm 0.9$ & $3.3 \pm 0.4$ & $518 \pm 339$ \\
45* & 150.0021 & +1.8278 & 24.8 & 26.4 & 26.0 & 26.2 & 24.5 & 25.9 & $5.8 \pm 0.6$ & $1.6 \pm 0.9$ & $2.7 \pm 0.4$ & $372 \pm 225$ \\
46* & 150.0638 & +1.4831 & 24.2 & 25.4 & 25.4 & 25.4 & 23.8 & 24.9 & $10.7 \pm 0.7$ & $4.0 \pm 1.0$ & $4.7 \pm 0.5$ & $265 \pm 75$ \\
47* & 150.0653 & +2.0156 & 24.4 & 26.2 & 25.7 & 26.0 & 24.1 & 25.2 & $8.6 \pm 0.5$ & $3.0 \pm 0.9$ & $3.8 \pm 0.4$ & $280 \pm 86$ \\
48* & 150.0710 & +2.7698 & 24.5 & 27.7 & 27.5 & 27.6 & 24.2 & 99.0 & $7.7 \pm 0.8$ & $<1.1$ & $4.2 \pm 0.5$ & $>831$ \\
49* & 150.0832 & +2.0176 & 25.1 & 27.1 & 26.8 & 27.0 & 24.8 & 28.3 & $4.2 \pm 0.6$ & $<0.9$ & $2.3 \pm 0.4$ & $>562$ \\
50* & 150.0937 & +2.6843 & 23.8 & 26.0 & 27.5 & 26.4 & 23.5 & 99.0 & $14.1 \pm 0.7$ & $<1.0$ & $7.9 \pm 0.5$ & $>1744$ \\
51* & 150.1005 & +2.7901 & 24.9 & 27.2 & 99.0 & 99.0 & 24.9 & 99.0 & $4.0 \pm 0.7$ & $<1.0$ & $2.7 \pm 0.5$ & $>592$ \\
52* & 150.1019 & +2.9165 & 24.6 & 26.2 & 25.5 & 25.8 & 24.2 & 25.0 & $7.3 \pm 0.7$ & $3.5 \pm 1.1$ & $3.0 \pm 0.5$ & $192 \pm 69$ \\
53* & 150.1090 & +1.5444 & 23.9 & 25.5 & 25.2 & 25.4 & 23.5 & 24.9 & $13.9 \pm 0.6$ & $4.1 \pm 0.9$ & $6.4 \pm 0.4$ & $352 \pm 84$ \\
54* & 150.1214 & +2.6877 & 25.1 & 25.9 & 99.0 & 99.0 & 24.5 & 26.9 & $5.8 \pm 0.7$ & $<1.0$ & $2.9 \pm 0.5$ & $>647$ \\
55* & 150.1267 & +2.2874 & 24.8 & 27.3 & 26.7 & 27.0 & 24.5 & 26.6 & $5.5 \pm 0.6$ & $<0.9$ & $2.8 \pm 0.4$ & $>649$ \\
56* & 150.1335 & +1.5006 & 25.0 & 26.6 & 25.8 & 26.1 & 24.6 & 25.9 & $5.1 \pm 0.6$ & $1.6 \pm 1.0$ & $2.3 \pm 0.4$ & $317 \pm 195$ \\
57* & 150.1376 & +2.2597 & 23.9 & 25.8 & 25.5 & 25.6 & 23.5 & 24.9 & $14.0 \pm 0.7$ & $4.2 \pm 1.0$ & $6.5 \pm 0.4$ & $344 \pm 83$ \\
58* & 150.1566 & +2.8614 & 25.0 & 28.9 & 27.7 & 28.2 & 24.8 & 99.0 & $4.2 \pm 0.7$ & $<1.0$ & $2.4 \pm 0.5$ & $>523$ \\
59* & 150.1676 & +2.3177 & 24.9 & 27.6 & 27.0 & 27.3 & 24.7 & 26.3 & $4.9 \pm 0.6$ & $1.1 \pm 1.0$ & $2.4 \pm 0.4$ & $455 \pm 413$ \\
60* & 150.1919 & +1.5765 & 24.5 & 25.8 & 25.8 & 25.8 & 24.1 & 25.5 & $8.3 \pm 0.7$ & $2.3 \pm 1.0$ & $3.9 \pm 0.5$ & $374 \pm 174$ \\
61* & 150.2032 & +2.2278 & 24.3 & 25.7 & 25.0 & 25.3 & 24.0 & 24.3 & $8.9 \pm 0.6$ & $6.7 \pm 0.9$ & $3.0 \pm 0.4$ & $100 \pm 19$ \\
62* & 150.2166 & +2.7730 & 24.8 & 27.5 & 99.0 & 99.0 & 24.5 & 99.0 & $5.7 \pm 0.7$ & $<1.1$ & $3.3 \pm 0.5$ & $>686$ \\
63* & 150.2254 & +1.5436 & 24.6 & 26.5 & 25.8 & 26.1 & 24.2 & 25.5 & $7.3 \pm 0.7$ & $2.3 \pm 1.0$ & $3.3 \pm 0.5$ & $320 \pm 151$ \\
64* & 150.2314 & +1.6086 & 24.9 & 27.6 & 26.6 & 27.0 & 24.5 & 25.8 & $5.8 \pm 0.6$ & $1.8 \pm 0.9$ & $2.6 \pm 0.4$ & $326 \pm 168$ \\
65* & 150.2434 & +1.6119 & 25.0 & 26.6 & 26.7 & 26.6 & 24.8 & 26.4 & $4.4 \pm 0.6$ & $1.0 \pm 0.9$ & $2.1 \pm 0.4$ & $456 \pm 413$ \\
66* & 150.2471 & +1.5555 & 24.6 & 26.5 & 24.8 & 25.4 & 24.3 & 24.5 & $6.7 \pm 0.7$ & $6.0 \pm 1.0$ & $2.0 \pm 0.5$ & $75 \pm 22$ \\
67* & 150.2521 & +2.8980 & 23.9 & 25.4 & 24.5 & 24.9 & 23.3 & 23.7 & $17.1 \pm 0.7$ & $12.6 \pm 1.2$ & $5.9 \pm 0.5$ & $104 \pm 13$ \\
68* & 150.2623 & +1.8624 & 24.9 & 26.9 & 26.8 & 26.9 & 24.5 & 26.7 & $5.8 \pm 0.6$ & $<0.9$ & $2.9 \pm 0.4$ & $>741$ \\
69* & 150.2807 & +1.8730 & 24.9 & 27.4 & 28.4 & 27.7 & 24.6 & 99.0 & $5.1 \pm 0.6$ & $<0.9$ & $2.8 \pm 0.4$ & $>679$ \\
70 & 150.2852 & +1.4858 & 24.5 & 27.0 & 28.2 & 27.4 & 24.3 & 27.5 & $6.8 \pm 0.8$ & $<1.3$ & $3.6 \pm 0.6$ & $>593$ \\
71* & 150.2905 & +2.2538 & 23.5 & 25.9 & 26.0 & 25.9 & 23.3 & 25.3 & $17.6 \pm 0.6$ & $2.8 \pm 0.9$ & $8.8 \pm 0.4$ & $704 \pm 230$ \\
72* & 150.2973 & +2.8944 & 25.0 & 27.3 & 26.7 & 27.0 & 24.8 & 25.8 & $4.3 \pm 0.7$ & $1.7 \pm 1.2$ & $1.8 \pm 0.5$ & $234 \pm 169$ \\
73* & 150.3267 & +1.9511 & 24.2 & 25.7 & 26.1 & 25.8 & 24.0 & 25.7 & $8.8 \pm 0.6$ & $1.9 \pm 1.0$ & $4.2 \pm 0.4$ & $488 \pm 250$ \\
74* & 150.3400 & +2.8002 & 24.9 & 26.6 & 26.0 & 26.3 & 24.4 & 25.6 & $6.2 \pm 0.7$ & $2.1 \pm 1.0$ & $2.8 \pm 0.5$ & $294 \pm 153$ \\
75* & 150.3493 & +1.9334 & 24.4 & 26.7 & 26.2 & 26.4 & 24.2 & 26.0 & $7.7 \pm 0.7$ & $1.4 \pm 1.0$ & $3.8 \pm 0.5$ & $602 \pm 442$ \\
76* & 150.3621 & +1.7417 & 24.3 & 28.0 & 25.9 & 26.6 & 24.0 & 25.6 & $8.8 \pm 0.6$ & $2.2 \pm 0.9$ & $4.2 \pm 0.4$ & $427 \pm 183$ \\
77* & 150.3657 & +2.5017 & 24.7 & 26.8 & 26.2 & 26.5 & 24.3 & 25.6 & $7.0 \pm 0.7$ & $2.0 \pm 1.0$ & $3.2 \pm 0.5$ & $355 \pm 184$ \\
78* & 150.3712 & +1.8250 & 23.9 & 26.1 & 25.4 & 25.8 & 23.5 & 25.0 & $14.2 \pm 0.6$ & $3.5 \pm 0.9$ & $6.7 \pm 0.4$ & $427 \pm 115$ \\
79* & 150.3795 & +2.5183 & 24.6 & 27.1 & 26.7 & 26.9 & 24.4 & 26.1 & $6.5 \pm 0.8$ & $1.3 \pm 1.0$ & $3.2 \pm 0.5$ & $530 \pm 427$ \\
80* & 150.3978 & +2.7734 & 24.8 & 26.5 & 26.4 & 26.4 & 24.3 & 25.5 & $6.6 \pm 0.7$ & $2.2 \pm 1.0$ & $3.0 \pm 0.5$ & $296 \pm 138$ \\
81* & 150.4005 & +1.8018 & 24.9 & 25.9 & 25.8 & 25.9 & 24.6 & 26.4 & $5.1 \pm 0.7$ & $<1.1$ & $2.5 \pm 0.5$ & $>510$ \\
82* & 150.4078 & +2.9117 & 24.8 & 26.4 & 25.8 & 26.1 & 24.6 & 25.4 & $5.2 \pm 0.7$ & $2.4 \pm 1.1$ & $2.2 \pm 0.5$ & $200 \pm 104$ \\
83* & 150.4079 & +2.1133 & 25.0 & 26.5 & 25.9 & 26.2 & 24.7 & 25.6 & $4.8 \pm 0.7$ & $2.1 \pm 0.9$ & $2.0 \pm 0.4$ & $210 \pm 103$ \\
84* & 150.4091 & +2.8063 & 24.1 & 25.3 & 24.5 & 24.9 & 23.5 & 24.0 & $13.9 \pm 0.6$ & $9.0 \pm 1.0$ & $5.1 \pm 0.4$ & $126 \pm 17$ \\
85* & 150.4274 & +2.4974 & 24.1 & 25.8 & 25.0 & 25.4 & 23.8 & 24.5 & $11.5 \pm 0.8$ & $5.8 \pm 1.0$ & $4.6 \pm 0.5$ & $178 \pm 37$ \\
86* & 150.4339 & +2.4867 & 25.0 & 27.1 & 27.0 & 27.0 & 24.8 & 26.3 & $4.6 \pm 0.7$ & $1.1 \pm 1.1$ & $2.2 \pm 0.5$ & $429 \pm 419$ \\
87* & 150.4393 & +2.7860 & 25.1 & 26.7 & 26.5 & 26.6 & 24.7 & 25.9 & $4.8 \pm 0.7$ & $1.6 \pm 1.0$ & $2.2 \pm 0.4$ & $306 \pm 199$ \\
88* & 150.4444 & +1.8076 & 24.7 & 27.5 & 99.0 & 99.0 & 24.5 & 99.0 & $5.6 \pm 0.7$ & $<1.0$ & $3.2 \pm 0.5$ & $>704$ \\
89* & 150.4685 & +2.7071 & 25.0 & 28.1 & 99.0 & 99.0 & 24.5 & 29.6 & $5.7 \pm 0.7$ & $<1.1$ & $3.1 \pm 0.5$ & $>619$ \\
90* & 150.4766 & +1.5314 & 25.0 & 26.7 & 99.0 & 99.0 & 24.7 & 99.0 & $4.7 \pm 0.7$ & $<1.1$ & $2.5 \pm 0.5$ & $>523$ \\
91* & 150.4890 & +1.6883 & 25.1 & 26.3 & 28.0 & 26.8 & 24.6 & 29.7 & $5.2 \pm 0.7$ & $<1.0$ & $2.8 \pm 0.5$ & $>595$ \\
92* & 150.4984 & +2.8138 & 25.0 & 27.5 & 99.0 & 99.0 & 24.7 & 99.0 & $5.0 \pm 0.8$ & $<1.1$ & $2.7 \pm 0.5$ & $>564$ \\
93* & 150.5114 & +2.7640 & 25.0 & 27.4 & 99.0 & 99.0 & 24.9 & 99.0 & $4.1 \pm 0.7$ & $<1.$ & $2.5 \pm 0.5$ & $>532$ \\
94* & 150.5131 & +1.6066 & 25.0 & 26.6 & 28.2 & 27.1 & 24.8 & 27.1 & $4.4 \pm 0.7$ & $<1.2$ & $2.3 \pm 0.5$ & $>431$ \\
95* & 150.5362 & +2.0874 & 24.9 & 27.0 & 27.2 & 27.1 & 24.7 & 26.9 & $4.8 \pm 0.7$ & $<1.0$ & $2.4 \pm 0.5$ & $>532$ \\
96* & 150.5367 & +1.9125 & 24.5 & 26.5 & 26.1 & 26.3 & 24.1 & 25.5 & $8.2 \pm 0.7$ & $2.2 \pm 1.0$ & $3.8 \pm 0.5$ & $378 \pm 167$ \\
97* & 150.5543 & +2.8230 & 24.7 & 26.8 & 26.1 & 26.4 & 24.4 & 25.6 & $6.4 \pm 0.7$ & $2.1 \pm 1.0$ & $2.9 \pm 0.5$ & $308 \pm 153$ \\
98* & 150.5677 & +2.5774 & 24.2 & 26.2 & 25.5 & 25.9 & 24.0 & 25.5 & $8.9 \pm 0.7$ & $2.4 \pm 1.0$ & $4.2 \pm 0.5$ & $388 \pm 175$ \\
99* & 150.5711 & +2.3625 & 24.9 & 26.1 & 26.3 & 26.2 & 24.6 & 26.1 & $5.2 \pm 0.6$ & $1.3 \pm 1.0$ & $2.5 \pm 0.4$ & $429 \pm 339$ \\
100* & 150.5773 & +1.6153 & 25.1 & 27.3 & 27.5 & 27.4 & 24.8 & 26.3 & $4.2 \pm 0.7$ & $1.1 \pm 1.1$ & $2.0 \pm 0.5$ & $392 \pm 389$ \\
101* & 150.6079 & +2.4935 & 25.0 & 28.2 & 99.0 & 99.0 & 24.8 & 99.0 & $4.4 \pm 0.7$ & $<1.0$ & $2.9 \pm 0.5$ & $>656$ \\
102* & 150.6386 & +2.3956 & 25.1 & 27.1 & 26.7 & 26.9 & 24.7 & 26.0 & $4.6 \pm 0.7$ & $1.4 \pm 1.1$ & $2.1 \pm 0.5$ & $337 \pm 273$ \\
103 & 150.6596 & +2.6453 & 24.7 & 27.5 & 26.6 & 27.1 & 24.4 & 26.1 & $6.5 \pm 0.7$ & $1.3 \pm 0.9$ & $3.2 \pm 0.5$ & $553 \pm 420$ \\
104 & 150.6806 & +2.7643 & 23.9 & 26.1 & 26.7 & 26.3 & 23.7 & 26.2 & $12.3 \pm 0.7$ & $1.2 \pm 0.9$ & $6.3 \pm 0.5$ & $1140 \pm 860$ \\
105* & 150.6927 & +1.8712 & 25.1 & 99.0 & 99.0 & 99.0 & 25.1 & 99.0 & $3.4 \pm 0.7$ & $<1.0$ & $1.8 \pm 0.5$ & $>398$ \\
106* & 150.7034 & +2.7397 & 25.0 & 27.5 & 99.0 & 99.0 & 24.7 & 99.0 & $4.9 \pm 0.7$ & $<1.0$ & $2.9 \pm 0.5$ & $>670$ \\
107* & 150.7111 & +2.2247 & 24.8 & 27.0 & 27.2 & 27.1 & 24.6 & 27.7 & $5.4 \pm 0.6$ & $<1.0$ & $2.8 \pm 0.4$ & $>614$ \\
108* & 150.7150 & +2.2342 & 24.4 & 26.8 & 26.8 & 26.8 & 24.2 & 26.1 & $7.7 \pm 0.6$ & $1.3 \pm 1.0$ & $3.8 \pm 0.4$ & $623 \pm 460$ \\
109* & 150.7475 & +2.8532 & 24.6 & 26.8 & 25.8 & 26.2 & 24.3 & 25.3 & $7.1 \pm 0.9$ & $2.8 \pm 1.2$ & $3.1 \pm 0.6$ & $246 \pm 113$ \\
110* & 150.7548 & +2.0434 & 24.6 & 26.6 & 25.1 & 25.7 & 24.3 & 24.7 & $6.7 \pm 0.8$ & $4.8 \pm 1.0$ & $2.4 \pm 0.5$ & $110 \pm 34$ \\
111* & 150.7576 & +1.8365 & 24.4 & 25.8 & 25.1 & 25.4 & 23.9 & 24.3 & $10.1 \pm 0.8$ & $7.0 \pm 1.2$ & $3.6 \pm 0.5$ & $114 \pm 25$ \\
112* & 150.7722 & +1.8614 & 25.1 & 27.3 & 26.5 & 26.9 & 24.6 & 26.0 & $5.3 \pm 0.7$ & $1.4 \pm 1.1$ & $2.5 \pm 0.5$ & $396 \pm 325$ \\
113* & 150.7747 & +2.1644 & 24.9 & 26.1 & 99.0 & 99.0 & 24.6 & 99.0 & $5.4 \pm 0.7$ & $<1.1$ & $3.0 \pm 0.5$ & $>617$ \\
114* & 150.7756 & +1.7953 & 24.3 & 27.0 & 99.0 & 99.0 & 24.0 & 99.0 & $9.5 \pm 0.7$ & $<1.1$ & $5.1 \pm 0.5$ & $>1040$ \\
115* & 150.7863 & +2.6448 & 23.7 & 25.4 & 25.2 & 25.3 & 23.4 & 24.6 & $16.2 \pm 0.8$ & $5.4 \pm 1.0$ & $7.3 \pm 0.5$ & $304 \pm 58$ \\
116* & 150.7906 & +2.2222 & 25.0 & 26.8 & 26.2 & 26.5 & 24.7 & 25.4 & $4.6 \pm 0.7$ & $2.6 \pm 1.0$ & $1.8 \pm 0.5$ & $151 \pm 71$ \\
117 & 150.8054 & +2.9250 & 24.5 & 25.8 & 25.1 & 25.4 & 24.1 & 24.8 & $8.1 \pm 1.0$ & $4.3 \pm 1.5$ & $3.2 \pm 0.7$ & $165 \pm 65$ \\
118* & 150.8211 & +2.2498 & 24.0 & 26.0 & 25.8 & 25.9 & 23.6 & 25.5 & $13.0 \pm 0.7$ & $2.3 \pm 1.1$ & $6.4 \pm 0.5$ & $631 \pm 304$ \\
119 & 150.8342 & +2.3398 & 25.0 & 27.0 & 26.4 & 26.7 & 24.5 & 25.2 & $5.8 \pm 0.8$ & $3.1 \pm 1.3$ & $2.3 \pm 0.6$ & $164 \pm 79$ \\
\enddata

\tablenotetext{a}{Asterisks denote the statistical sample.}
\tablenotetext{b}{AB magnitude.
 An entry of ``99.0'' indicates that no excess flux was measured.
 All of our LAE candidates are undetected in the $B$-, $g^\prime$-, $V$-, $r^\prime$-band data.
}
\tablenotetext{c}{Flux densities of the {\it NB816} band and the $z^\prime$ band
in unit of $10^{-30}$ erg s$^{-1}$ cm$^{-2}$ Hz$^{-1}$.
Errors and upper-limits represent 1$\sigma$ significance.}
\tablenotetext{d}{Estimated line fluxes in unit of $10^{-17}$ erg s$^{-1}$ cm$^{-2}$.  Errors represent 1$\sigma$ significance.}
\tablenotetext{e}{Estimated observed equivalent widths. Errors and lower-limits represent 1$\sigma$ significance.}
\end{deluxetable}

\clearpage
\LongTables
\begin{deluxetable}{ccccccc}
\tablenum{2}
\tabletypesize{\scriptsize}
\tablecaption{Ly$\alpha$ luminosity and star formation rate for the
LAE candidates at $z \approx 5.7$\label{tab:LLAE}}
\tablewidth{0pt}
\tablehead{
\colhead{No.\tablenotemark{a}} &
\colhead{$EW_0({\rm Ly}\alpha)$\tablenotemark{b}} &
\colhead{$L({\rm Ly}\alpha)$\tablenotemark{b}} &
\colhead{$SFR$(Ly$\alpha$)\tablenotemark{b}}  &
\colhead{$L_{1270}$\tablenotemark{b,c}} &
\colhead{$SFR$(UV)\tablenotemark{b}} &
\colhead{$SFR({\rm Ly\alpha})/SFR({\rm UV})$\tablenotemark{b}} \\
\colhead{} &
\colhead{(\AA)} &
\colhead{($10^{42}$ ergs s$^{-1}$)} &
\colhead{($M_\odot$ yr$^{-1}$)} &
\colhead{($10^{28}$ ergs s$^{-1}$ Hz$^{-1}$)} &
\colhead{($M_\odot$ yr$^{-1}$)} &
\colhead{} \\
}

\startdata
1* & $53 \pm 31$ & $9.2 \pm 1.4$ & $8.4 \pm 1.3$ & $8.7 \pm 4.9$ & $12.1 \pm 6.8$ & $0.69 \pm 0.41$ \\
2* & $46 \pm 29$ & $8.0 \pm 1.6$ & $7.3 \pm 1.5$ & $8.7 \pm 5.1$ & $12.1 \pm 7.2$ & $0.60 \pm 0.38$ \\
3* & $68 \pm 60$ & $8.2 \pm 1.6$ & $7.4 \pm 1.5$ & $6.0 \pm 5.2$ & $8.4 \pm 7.2$ & $0.89 \pm 0.79$ \\
4* & $67 \pm 16$ & $31.0 \pm 1.6$ & $28.2 \pm 1.4$ & $22.7 \pm 5.2$ & $31.8 \pm 7.3$ & $0.89 \pm 0.21$ \\
5* & $86 \pm 74$ & $10.7 \pm 1.7$ & $9.7 \pm 1.6$ & $6.1 \pm 5.2$ & $8.6 \pm 7.3$ & $1.14 \pm 0.98$ \\
6* &$>74$ & $8.1 \pm 1.7$ & $7.4 \pm 1.6$ & $<5.4$ & $<7.5$ & $>0.98$ \\
7* & $34 \pm 16$ & $7.9 \pm 1.6$ & $7.2 \pm 1.4$ & $11.5 \pm 4.8$ & $16.1 \pm 6.8$ & $0.45 \pm 0.21$ \\
8* & $20 \pm 6$ & $9.8 \pm 1.7$ & $9.0 \pm 1.6$ & $24.0 \pm 5.5$ & $33.6 \pm 7.7$ & $0.27 \pm 0.08$ \\
9* &$>98$ & $10.4 \pm 1.7$ & $9.4 \pm 1.6$ & $<5.2$ & $<7.3$ & $>1.29$ \\
10* & $22 \pm 6$ & $11.8 \pm 1.8$ & $10.8 \pm 1.6$ & $26.6 \pm 5.8$ & $37.2 \pm 8.1$ & $0.29 \pm 0.08$ \\
11 & $25 \pm 9$ & $12.2 \pm 2.2$ & $11.1 \pm 2.0$ & $23.6 \pm 7.2$ & $33.1 \pm 10.1$ & $0.34 \pm 0.12$ \\
12* &$>69$ & $8.8 \pm 1.9$ & $8.0 \pm 1.7$ & $<6.3$ & $<8.8$ & $>0.91$ \\
13* &$>175$ & $18.4 \pm 1.6$ & $16.8 \pm 1.5$ & $<5.2$ & $<7.3$ & $>2.30$ \\
14* & $55 \pm 16$ & $21.2 \pm 1.7$ & $19.3 \pm 1.5$ & $19.0 \pm 5.2$ & $26.6 \pm 7.3$ & $0.72 \pm 0.21$ \\
15* & $19 \pm 7$ & $6.7 \pm 1.6$ & $6.1 \pm 1.5$ & $17.4 \pm 4.9$ & $24.3 \pm 6.9$ & $0.25 \pm 0.09$ \\
16* &$>68$ & $7.1 \pm 1.7$ & $6.4 \pm 1.5$ & $<5.1$ & $<7.2$ & $>0.89$ \\
17* & $35 \pm 14$ & $11.7 \pm 2.0$ & $10.6 \pm 1.8$ & $16.3 \pm 6.0$ & $22.8 \pm 8.3$ & $0.47 \pm 0.19$ \\
18* & $91 \pm 57$ & $19.1 \pm 2.1$ & $17.4 \pm 1.9$ & $10.3 \pm 6.4$ & $14.4 \pm 8.9$ & $1.20 \pm 0.76$ \\
19* & $77 \pm 54$ & $13.0 \pm 1.7$ & $11.9 \pm 1.5$ & $8.3 \pm 5.7$ & $11.7 \pm 8.0$ & $1.02 \pm 0.71$ \\
20* & $51 \pm 36$ & $8.0 \pm 1.5$ & $7.2 \pm 1.4$ & $7.7 \pm 5.2$ & $10.7 \pm 7.2$ & $0.68 \pm 0.47$ \\
21* &$>79$ & $8.4 \pm 1.8$ & $7.6 \pm 1.6$ & $<5.2$ & $<7.3$ & $>1.04$ \\
22 &$>74$ & $12.8 \pm 2.1$ & $11.7 \pm 1.9$ & $<8.5$ & $<11.9$ & $>0.98$ \\
23* & $58 \pm 35$ & $10.1 \pm 1.6$ & $9.2 \pm 1.5$ & $8.6 \pm 5.1$ & $12.0 \pm 7.1$ & $0.76 \pm 0.47$ \\
24* &$>100$ & $12.3 \pm 1.7$ & $11.2 \pm 1.6$ & $<6.1$ & $<8.5$ & $>1.32$ \\
25* &$>94$ & $9.3 \pm 1.4$ & $8.5 \pm 1.3$ & $<4.9$ & $<6.8$ & $>1.24$ \\
26* &$>71$ & $6.5 \pm 1.5$ & $5.9 \pm 1.3$ & $<4.5$ & $<6.4$ & $>0.93$ \\
27* & $85 \pm 36$ & $16.3 \pm 1.3$ & $14.8 \pm 1.1$ & $9.4 \pm 3.9$ & $13.2 \pm 5.4$ & $1.12 \pm 0.47$ \\
28* &$>77$ & $7.6 \pm 1.5$ & $7.0 \pm 1.4$ & $<4.9$ & $<6.8$ & $>1.02$ \\
29* & $72 \pm 49$ & $11.1 \pm 1.4$ & $10.1 \pm 1.3$ & $7.6 \pm 5.1$ & $10.7 \pm 7.1$ & $0.95 \pm 0.64$ \\
30* & $66 \pm 43$ & $12.2 \pm 1.7$ & $11.1 \pm 1.5$ & $9.1 \pm 5.7$ & $12.7 \pm 8.0$ & $0.87 \pm 0.56$ \\
31* & $154 \pm 114$ & $20.3 \pm 1.5$ & $18.5 \pm 1.4$ & $6.5 \pm 4.8$ & $9.1 \pm 6.7$ & $2.03 \pm 1.51$ \\
32 & $127 \pm 112$ & $18.5 \pm 2.1$ & $16.8 \pm 2.0$ & $7.2 \pm 6.3$ & $10.1 \pm 8.8$ & $1.67 \pm 1.47$ \\
33* & $30 \pm 10$ & $11.0 \pm 1.8$ & $10.0 \pm 1.6$ & $18.3 \pm 5.3$ & $25.6 \pm 7.5$ & $0.39 \pm 0.13$ \\
34* & $29 \pm 13$ & $7.3 \pm 1.5$ & $6.6 \pm 1.3$ & $12.6 \pm 5.2$ & $17.6 \pm 7.3$ & $0.38 \pm 0.17$ \\
35* & $42 \pm 18$ & $9.6 \pm 1.3$ & $8.7 \pm 1.2$ & $11.2 \pm 4.5$ & $15.7 \pm 6.2$ & $0.56 \pm 0.24$ \\
36* & $56 \pm 36$ & $9.4 \pm 1.6$ & $8.5 \pm 1.5$ & $8.3 \pm 5.1$ & $11.6 \pm 7.2$ & $0.74 \pm 0.47$ \\
37* &$>100$ & $12.4 \pm 1.8$ & $11.3 \pm 1.6$ & $<6.1$ & $<8.6$ & $>1.31$ \\
38* &$>100$ & $10.9 \pm 1.7$ & $9.9 \pm 1.5$ & $<5.4$ & $<7.5$ & $>1.32$ \\
39* & $89 \pm 56$ & $13.6 \pm 1.4$ & $12.3 \pm 1.3$ & $7.5 \pm 4.7$ & $10.5 \pm 6.6$ & $1.17 \pm 0.74$ \\
40* & $20 \pm 4$ & $13.3 \pm 1.5$ & $12.1 \pm 1.4$ & $32.3 \pm 5.1$ & $45.2 \pm 7.1$ & $0.27 \pm 0.05$ \\
41* &$>103$ & $11.2 \pm 1.7$ & $10.2 \pm 1.5$ & $<5.3$ & $<7.5$ & $>1.36$ \\
42* &$>72$ & $8.0 \pm 1.6$ & $7.3 \pm 1.4$ & $<5.5$ & $<7.8$ & $>0.94$ \\
43* & $91 \pm 65$ & $13.2 \pm 1.5$ & $12.0 \pm 1.3$ & $7.2 \pm 5.1$ & $10.0 \pm 7.1$ & $1.20 \pm 0.86$ \\
44* & $77 \pm 51$ & $11.6 \pm 1.5$ & $10.6 \pm 1.4$ & $7.4 \pm 4.8$ & $10.4 \pm 6.7$ & $1.02 \pm 0.67$ \\
45* & $55 \pm 34$ & $9.6 \pm 1.5$ & $8.8 \pm 1.3$ & $8.6 \pm 5.0$ & $12.0 \pm 7.0$ & $0.73 \pm 0.44$ \\
46* & $39 \pm 11$ & $16.7 \pm 1.7$ & $15.2 \pm 1.6$ & $20.9 \pm 5.5$ & $29.3 \pm 7.7$ & $0.52 \pm 0.15$ \\
47* & $42 \pm 13$ & $13.5 \pm 1.3$ & $12.3 \pm 1.2$ & $15.9 \pm 4.6$ & $22.3 \pm 6.5$ & $0.55 \pm 0.17$ \\
48* &$>124$ & $14.8 \pm 1.8$ & $13.5 \pm 1.6$ & $<5.9$ & $<8.2$ & $>1.63$ \\
49* &$>84$ & $8.0 \pm 1.4$ & $7.3 \pm 1.2$ & $<4.7$ & $<6.6$ & $>1.11$ \\
50* &$>260$ & $27.9 \pm 1.6$ & $25.4 \pm 1.5$ & $<5.3$ & $<7.4$ & $>3.43$ \\
51* &$>88$ & $9.5 \pm 1.7$ & $8.7 \pm 1.5$ & $<5.3$ & $<7.4$ & $>1.16$ \\
52* & $29 \pm 10$ & $10.7 \pm 1.8$ & $9.7 \pm 1.6$ & $18.4 \pm 5.9$ & $25.8 \pm 8.2$ & $0.38 \pm 0.14$ \\
53* & $53 \pm 13$ & $22.8 \pm 1.5$ & $20.7 \pm 1.4$ & $21.4 \pm 4.9$ & $30.0 \pm 6.9$ & $0.69 \pm 0.17$ \\
54* &$>97$ & $10.4 \pm 1.7$ & $9.5 \pm 1.5$ & $<5.3$ & $<7.5$ & $>1.27$ \\
55* &$>97$ & $9.8 \pm 1.5$ & $8.9 \pm 1.3$ & $<5.0$ & $<7.0$ & $>1.28$ \\
56* & $47 \pm 29$ & $8.2 \pm 1.5$ & $7.5 \pm 1.4$ & $8.6 \pm 5.0$ & $12.0 \pm 7.1$ & $0.62 \pm 0.38$ \\
57* & $51 \pm 12$ & $22.8 \pm 1.6$ & $20.8 \pm 1.4$ & $22.0 \pm 5.1$ & $30.8 \pm 7.2$ & $0.68 \pm 0.16$ \\
58* &$>78$ & $8.5 \pm 1.7$ & $7.8 \pm 1.5$ & $<5.4$ & $<7.6$ & $>1.03$ \\
59* & $68 \pm 62$ & $8.3 \pm 1.5$ & $7.6 \pm 1.4$ & $6.1 \pm 5.4$ & $8.5 \pm 7.6$ & $0.89 \pm 0.81$ \\
60* & $56 \pm 26$ & $13.7 \pm 1.7$ & $12.5 \pm 1.5$ & $12.1 \pm 5.5$ & $17.0 \pm 7.6$ & $0.73 \pm 0.34$ \\
61* & $15 \pm 3$ & $10.7 \pm 1.5$ & $9.7 \pm 1.3$ & $35.2 \pm 4.8$ & $49.3 \pm 6.7$ & $0.20 \pm 0.04$ \\
62* &$>102$ & $11.5 \pm 1.7$ & $10.5 \pm 1.6$ & $<5.6$ & $<7.8$ & $>1.35$ \\
63* & $48 \pm 23$ & $11.8 \pm 1.6$ & $10.8 \pm 1.5$ & $12.2 \pm 5.5$ & $17.1 \pm 7.7$ & $0.63 \pm 0.30$ \\
64* & $49 \pm 25$ & $9.4 \pm 1.4$ & $8.5 \pm 1.3$ & $9.5 \pm 4.7$ & $13.3 \pm 6.6$ & $0.64 \pm 0.33$ \\
65* & $68 \pm 62$ & $7.4 \pm 1.4$ & $6.7 \pm 1.3$ & $5.3 \pm 4.7$ & $7.5 \pm 6.6$ & $0.90 \pm 0.81$ \\
66* & $11 \pm 3$ & $7.1 \pm 1.7$ & $6.5 \pm 1.5$ & $31.5 \pm 5.3$ & $44.1 \pm 7.5$ & $0.15 \pm 0.04$ \\
67* & $15 \pm 2$ & $20.8 \pm 1.8$ & $18.9 \pm 1.6$ & $66.3 \pm 6.3$ & $92.9 \pm 8.9$ & $0.20 \pm 0.03$ \\
68* &$>111$ & $10.4 \pm 1.4$ & $9.4 \pm 1.3$ & $<4.6$ & $<6.5$ & $>1.46$ \\
69* &$>101$ & $10.0 \pm 1.4$ & $9.1 \pm 1.3$ & $<4.9$ & $<6.8$ & $>1.33$ \\
70 &$>89$ & $12.6 \pm 2.0$ & $11.5 \pm 1.8$ & $<7.0$ & $<9.8$ & $>1.17$ \\
71* & $105 \pm 34$ & $31.1 \pm 1.4$ & $28.3 \pm 1.3$ & $14.6 \pm 4.7$ & $20.5 \pm 6.6$ & $1.38 \pm 0.45$ \\
72* & $35 \pm 25$ & $6.5 \pm 1.8$ & $5.9 \pm 1.6$ & $9.2 \pm 6.2$ & $12.9 \pm 8.7$ & $0.46 \pm 0.33$ \\
73* & $73 \pm 37$ & $15.0 \pm 1.5$ & $13.7 \pm 1.4$ & $10.2 \pm 5.1$ & $14.2 \pm 7.2$ & $0.96 \pm 0.49$ \\
74* & $44 \pm 23$ & $9.9 \pm 1.7$ & $9.0 \pm 1.6$ & $11.1 \pm 5.4$ & $15.5 \pm 7.6$ & $0.58 \pm 0.30$ \\
75* & $90 \pm 66$ & $13.4 \pm 1.6$ & $12.2 \pm 1.5$ & $7.4 \pm 5.3$ & $10.3 \pm 7.4$ & $1.18 \pm 0.87$ \\
76* & $64 \pm 27$ & $14.8 \pm 1.4$ & $13.5 \pm 1.3$ & $11.5 \pm 4.8$ & $16.0 \pm 6.7$ & $0.84 \pm 0.36$ \\
77* & $53 \pm 27$ & $11.5 \pm 1.6$ & $10.5 \pm 1.5$ & $10.7 \pm 5.3$ & $15.0 \pm 7.4$ & $0.70 \pm 0.36$ \\
78* & $64 \pm 17$ & $23.8 \pm 1.4$ & $21.7 \pm 1.3$ & $18.4 \pm 4.8$ & $25.8 \pm 6.8$ & $0.84 \pm 0.23$ \\
79* & $79 \pm 64$ & $11.2 \pm 1.9$ & $10.2 \pm 1.7$ & $7.0 \pm 5.5$ & $9.8 \pm 7.7$ & $1.04 \pm 0.84$ \\
80* & $44 \pm 21$ & $10.6 \pm 1.7$ & $9.6 \pm 1.5$ & $11.8 \pm 5.2$ & $16.5 \pm 7.2$ & $0.58 \pm 0.27$ \\
81* &$>76$ & $8.9 \pm 1.7$ & $8.1 \pm 1.6$ & $<5.7$ & $<8.0$ & $>1.00$ \\
82* & $30 \pm 15$ & $7.7 \pm 1.8$ & $7.0 \pm 1.6$ & $12.7 \pm 5.9$ & $17.8 \pm 8.3$ & $0.39 \pm 0.20$ \\
83* & $31 \pm 15$ & $7.1 \pm 1.6$ & $6.5 \pm 1.4$ & $11.2 \pm 4.9$ & $15.7 \pm 6.9$ & $0.41 \pm 0.20$ \\
84* & $19 \pm 3$ & $18.1 \pm 1.5$ & $16.5 \pm 1.4$ & $47.5 \pm 5.1$ & $66.5 \pm 7.1$ & $0.25 \pm 0.03$ \\
85* & $27 \pm 5$ & $16.4 \pm 1.8$ & $14.9 \pm 1.6$ & $30.5 \pm 5.3$ & $42.7 \pm 7.4$ & $0.35 \pm 0.07$ \\
86* & $64 \pm 63$ & $7.7 \pm 1.6$ & $7.0 \pm 1.5$ & $5.9 \pm 5.6$ & $8.3 \pm 7.9$ & $0.84 \pm 0.82$ \\
87* & $46 \pm 30$ & $7.8 \pm 1.6$ & $7.1 \pm 1.4$ & $8.4 \pm 5.2$ & $11.7 \pm 7.3$ & $0.60 \pm 0.39$ \\
88* &$>105$ & $11.4 \pm 1.7$ & $10.4 \pm 1.5$ & $<5.3$ & $<7.5$ & $>1.38$ \\
89* &$>92$ & $10.8 \pm 1.8$ & $9.8 \pm 1.6$ & $<5.8$ & $<8.1$ & $>1.22$ \\
90* &$>78$ & $9.0 \pm 1.7$ & $8.2 \pm 1.5$ & $<5.7$ & $<8.0$ & $>1.03$ \\
91* &$>89$ & $9.9 \pm 1.7$ & $9.0 \pm 1.5$ & $<5.5$ & $<7.7$ & $>1.17$ \\
92* &$>84$ & $9.5 \pm 1.8$ & $8.7 \pm 1.6$ & $<5.6$ & $<7.8$ & $>1.11$ \\
93* &$>79$ & $8.8 \pm 1.7$ & $8.0 \pm 1.5$ & $<5.5$ & $<7.7$ & $>1.05$ \\
94* &$>64$ & $8.0 \pm 1.7$ & $7.3 \pm 1.6$ & $<6.1$ & $<8.6$ & $>0.85$ \\
95* &$>79$ & $8.5 \pm 1.6$ & $7.7 \pm 1.5$ & $<5.3$ & $<7.4$ & $>1.05$ \\
96* & $56 \pm 25$ & $13.5 \pm 1.7$ & $12.3 \pm 1.5$ & $11.8 \pm 5.0$ & $16.5 \pm 7.0$ & $0.74 \pm 0.33$ \\
97* & $46 \pm 23$ & $10.2 \pm 1.6$ & $9.3 \pm 1.5$ & $10.9 \pm 5.1$ & $15.3 \pm 7.2$ & $0.61 \pm 0.30$ \\
98* & $58 \pm 26$ & $14.8 \pm 1.7$ & $13.4 \pm 1.6$ & $12.6 \pm 5.5$ & $17.6 \pm 7.7$ & $0.76 \pm 0.34$ \\
99* & $64 \pm 51$ & $8.7 \pm 1.5$ & $7.9 \pm 1.4$ & $6.7 \pm 5.2$ & $9.4 \pm 7.3$ & $0.84 \pm 0.67$ \\
100* & $58 \pm 58$ & $7.1 \pm 1.7$ & $6.4 \pm 1.6$ & $6.0 \pm 5.7$ & $8.3 \pm 8.0$ & $0.77 \pm 0.76$ \\
101* &$>98$ & $10.2 \pm 1.7$ & $9.3 \pm 1.5$ & $<5.1$ & $<7.2$ & $>1.29$ \\
102* & $50 \pm 41$ & $7.6 \pm 1.7$ & $6.9 \pm 1.6$ & $7.4 \pm 5.7$ & $10.4 \pm 8.0$ & $0.66 \pm 0.54$ \\
103 & $83 \pm 63$ & $11.2 \pm 1.6$ & $10.2 \pm 1.5$ & $6.7 \pm 5.0$ & $9.3 \pm 6.9$ & $1.09 \pm 0.82$ \\
104 & $170 \pm 128$ & $22.3 \pm 1.7$ & $20.3 \pm 1.5$ & $6.5 \pm 4.9$ & $9.1 \pm 6.8$ & $2.24 \pm 1.69$ \\
105* &$>59$ & $6.5 \pm 1.6$ & $5.9 \pm 1.5$ & $<5.4$ & $<7.6$ & $>0.78$ \\
106* &$>100$ & $10.2 \pm 1.6$ & $9.3 \pm 1.5$ & $<5.0$ & $<7.1$ & $>1.32$ \\
107* &$>92$ & $10.1 \pm 1.6$ & $9.2 \pm 1.4$ & $<5.4$ & $<7.6$ & $>1.21$ \\
108* & $93 \pm 69$ & $13.4 \pm 1.5$ & $12.2 \pm 1.4$ & $7.1 \pm 5.2$ & $9.9 \pm 7.2$ & $1.23 \pm 0.90$ \\
109* & $37 \pm 17$ & $10.9 \pm 2.0$ & $9.9 \pm 1.8$ & $14.6 \pm 6.1$ & $20.4 \pm 8.6$ & $0.48 \pm 0.22$ \\
110* & $16 \pm 5$ & $8.4 \pm 1.9$ & $7.6 \pm 1.7$ & $25.1 \pm 5.4$ & $35.2 \pm 7.6$ & $0.22 \pm 0.07$ \\
111* & $17 \pm 4$ & $12.6 \pm 1.9$ & $11.5 \pm 1.7$ & $36.8 \pm 6.1$ & $51.5 \pm 8.5$ & $0.22 \pm 0.05$ \\
112* & $59 \pm 48$ & $8.8 \pm 1.7$ & $8.0 \pm 1.6$ & $7.4 \pm 5.9$ & $10.3 \pm 8.2$ & $0.78 \pm 0.64$ \\
113* &$>92$ & $10.5 \pm 1.7$ & $9.6 \pm 1.5$ & $<5.6$ & $<7.9$ & $>1.21$ \\
114* &$>155$ & $18.1 \pm 1.8$ & $16.5 \pm 1.6$ & $<5.8$ & $<8.1$ & $>2.04$ \\
115* & $45 \pm 9$ & $26.0 \pm 1.7$ & $23.6 \pm 1.6$ & $28.3 \pm 5.1$ & $39.6 \pm 7.1$ & $0.60 \pm 0.11$ \\
116* & $23 \pm 11$ & $6.3 \pm 1.6$ & $5.7 \pm 1.5$ & $13.8 \pm 5.4$ & $19.3 \pm 7.6$ & $0.30 \pm 0.14$ \\
117 & $25 \pm 10$ & $11.4 \pm 2.3$ & $10.4 \pm 2.1$ & $22.9 \pm 7.7$ & $32.1 \pm 10.8$ & $0.32 \pm 0.13$ \\
118* & $94 \pm 45$ & $22.8 \pm 1.7$ & $20.7 \pm 1.5$ & $11.9 \pm 5.7$ & $16.7 \pm 8.0$ & $1.24 \pm 0.60$ \\
119 & $25 \pm 12$ & $8.2 \pm 2.0$ & $7.5 \pm 1.8$ & $16.5 \pm 6.9$ & $23.1 \pm 9.6$ & $0.32 \pm 0.16$ \\
\enddata
\tablenotetext{a}{Asterisks denote the statistical sample.}
\tablenotetext{b}{Errors and lower-limits represent 1$\sigma$ significance.}
\tablenotetext{c}{The UV continuum luminosity at $\lambda=1270$ \AA.}
\end{deluxetable}

\clearpage


\clearpage

%

\begin{figure}
\plotone{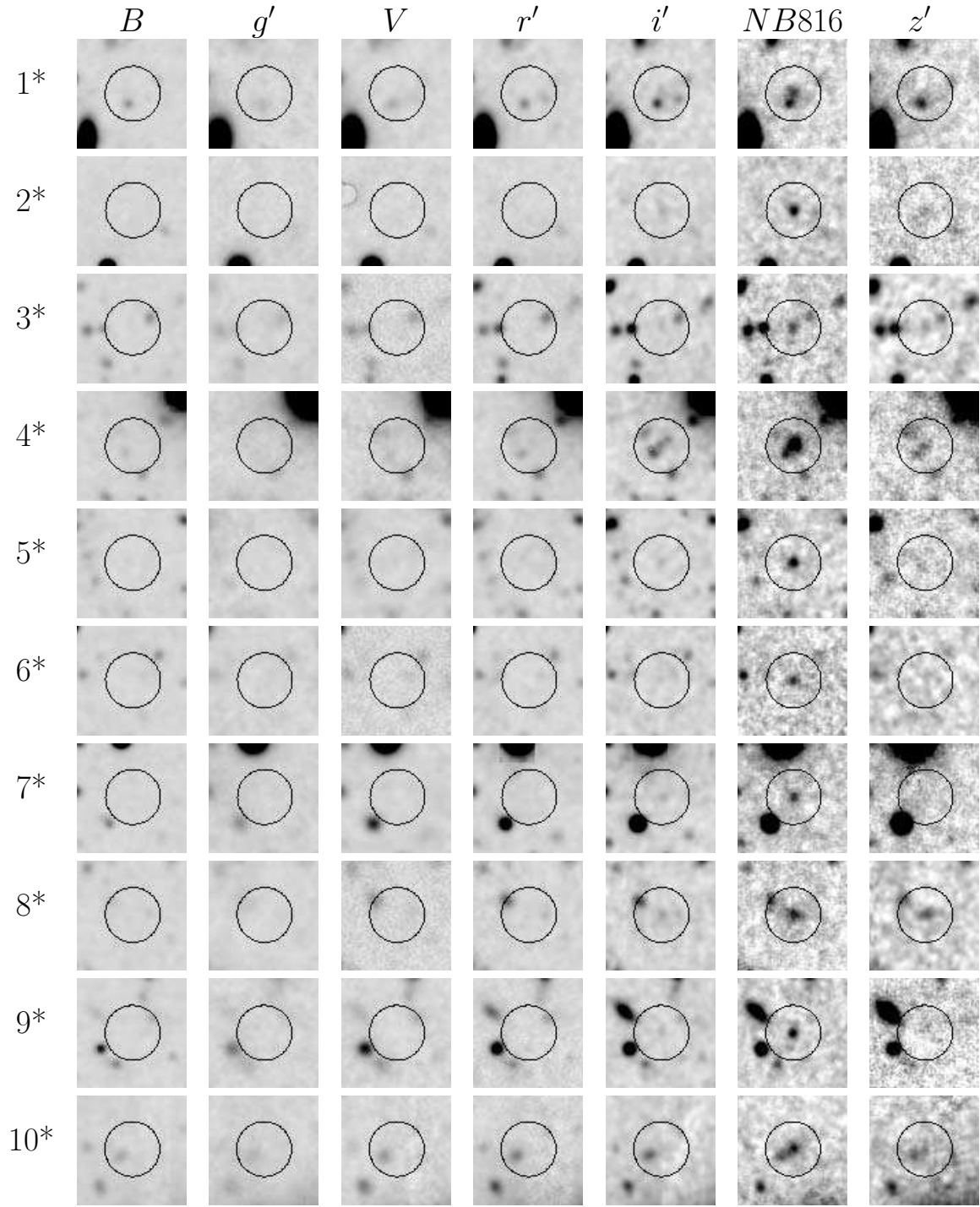}
\caption{Broad-band and \NB ~ images of our
          LAE candidates at $z \approx 5.7$.
         Asterisks at the object number denote the statistical sample.
         Each box is $12^{\prime \prime}$ on a side
         (north is up and east is left).
         Each circle has $3^{\prime \prime}$ radius.\label{thum}
}
\end{figure}
\clearpage
{\plotone{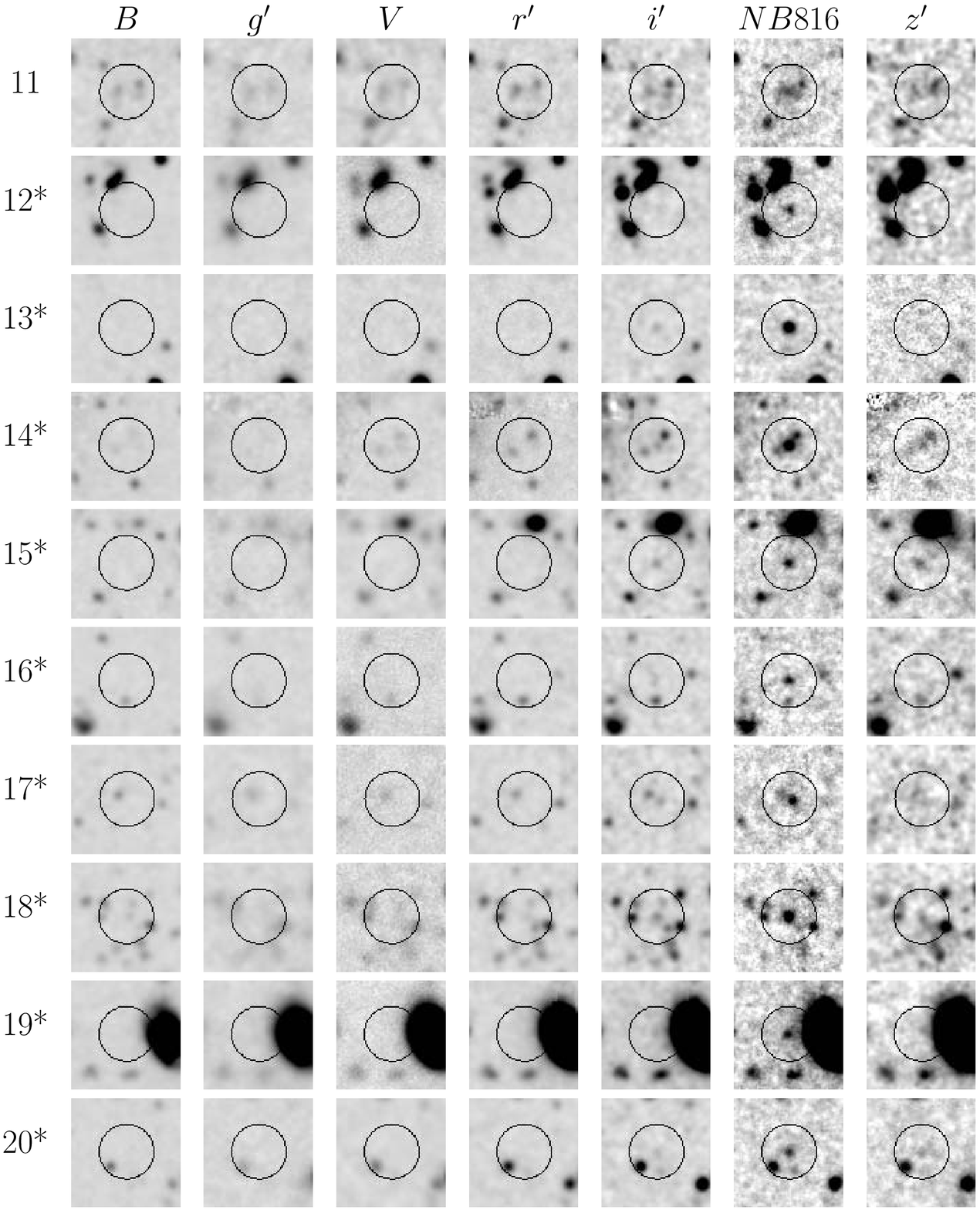}}\\[2mm]
\centerline{Fig. 2. --- continued.}
\clearpage
{\plotone{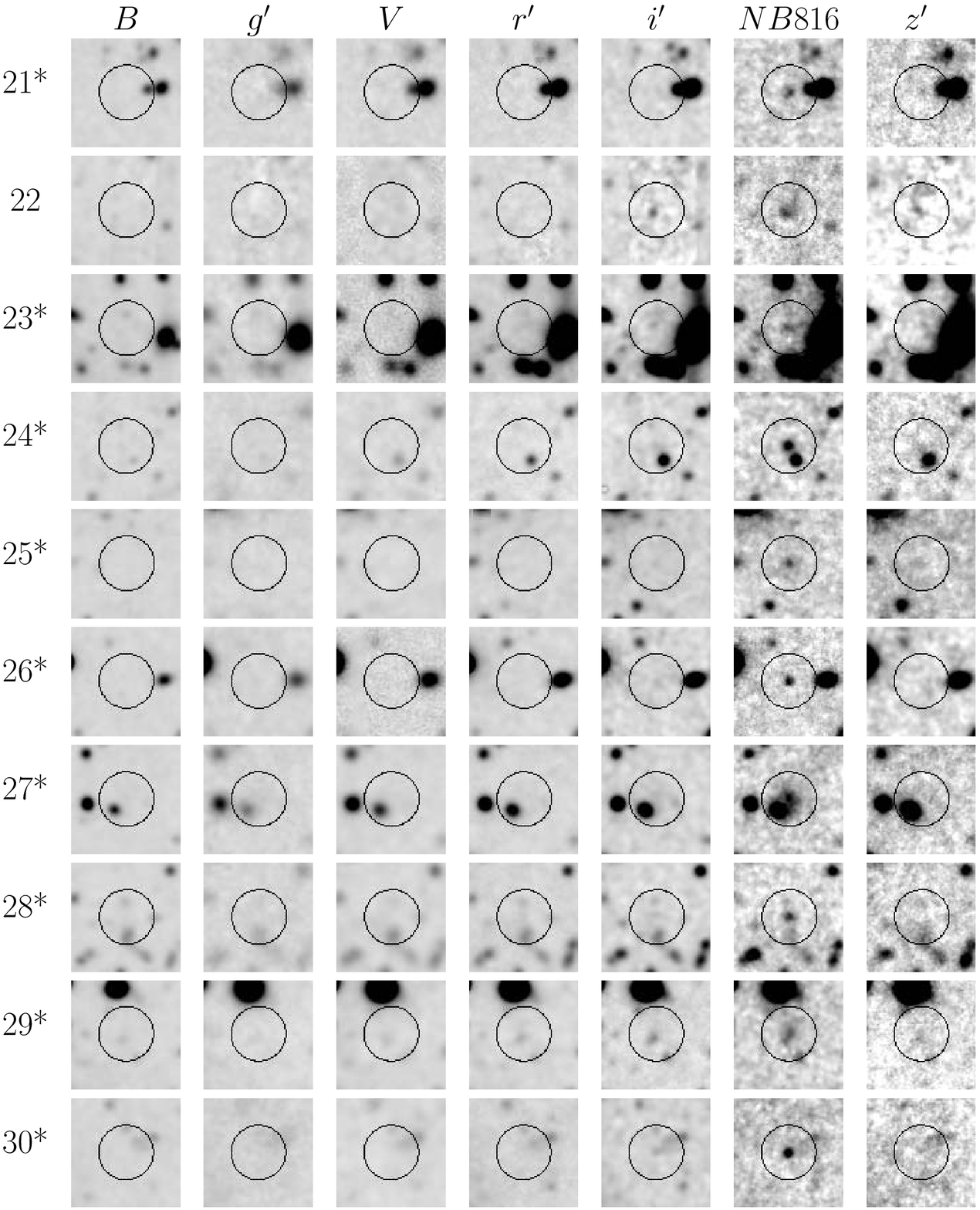}}\\[2mm]
\centerline{Fig. 2. --- continued.}
\clearpage
{\plotone{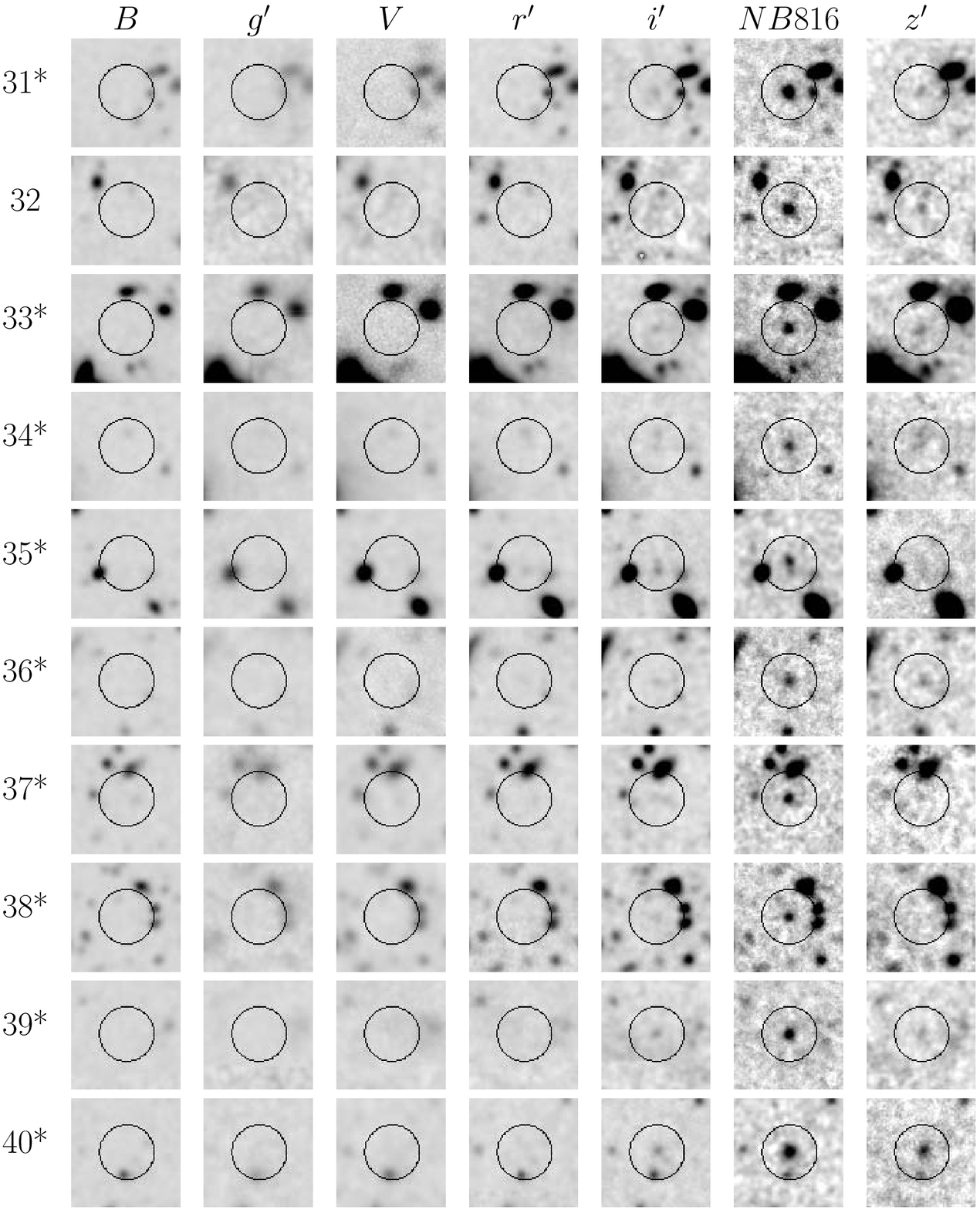}}\\[2mm]
\centerline{Fig. 2. --- continued.}
\clearpage
{\plotone{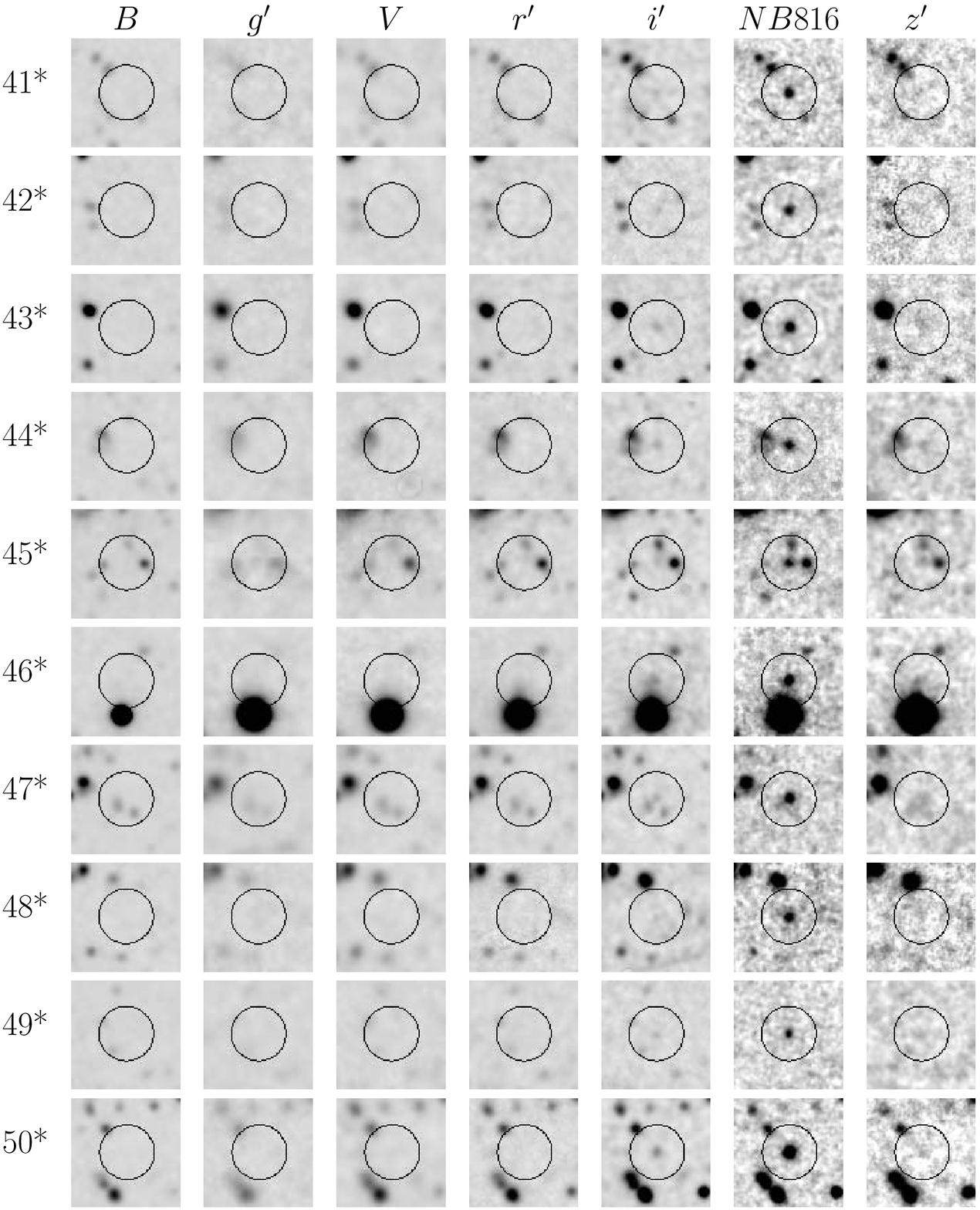}}\\[2mm]
\centerline{Fig. 2. --- continued.}
\clearpage
{\plotone{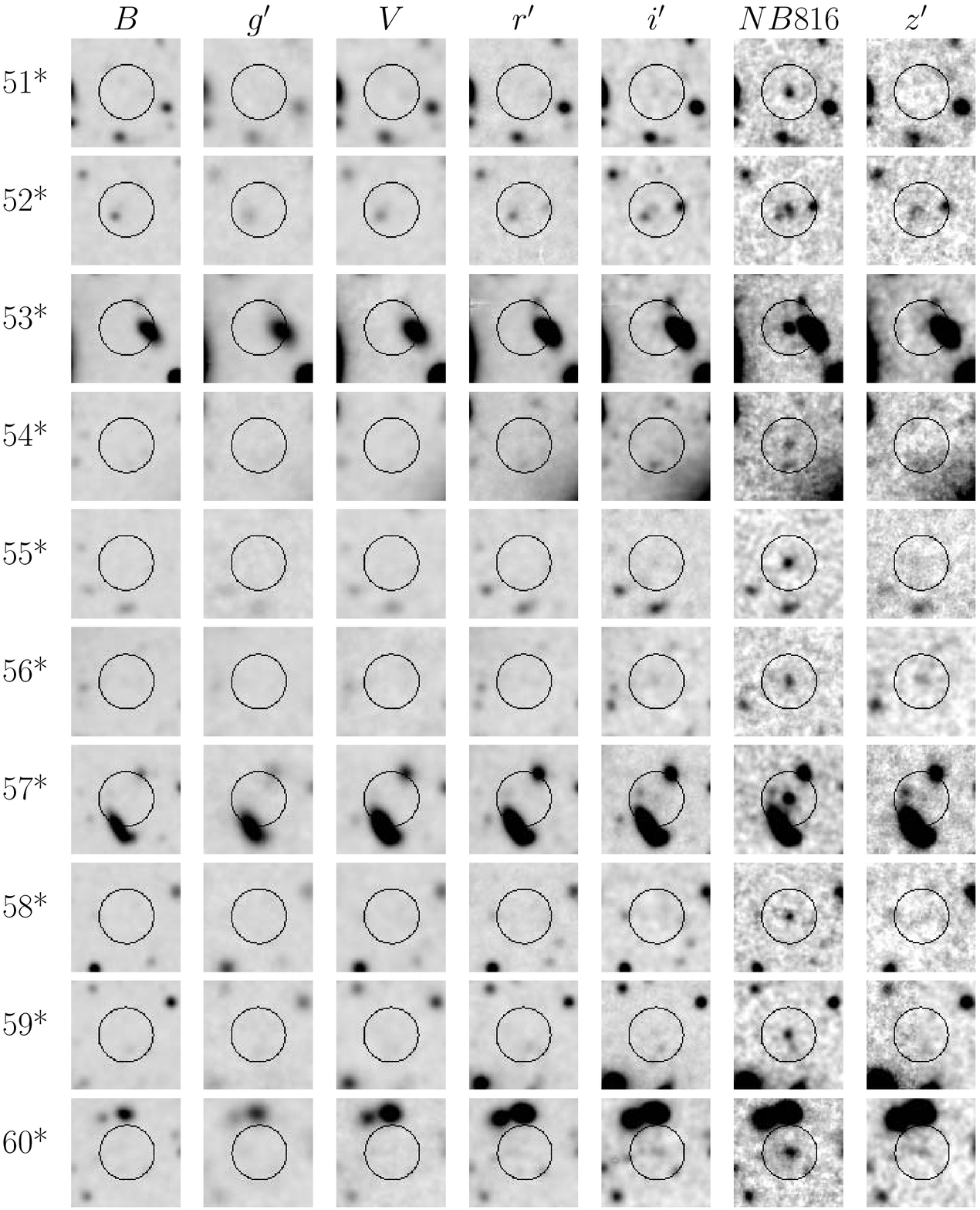}}\\[2mm]
\centerline{Fig. 2. --- continued.}
\clearpage
{\plotone{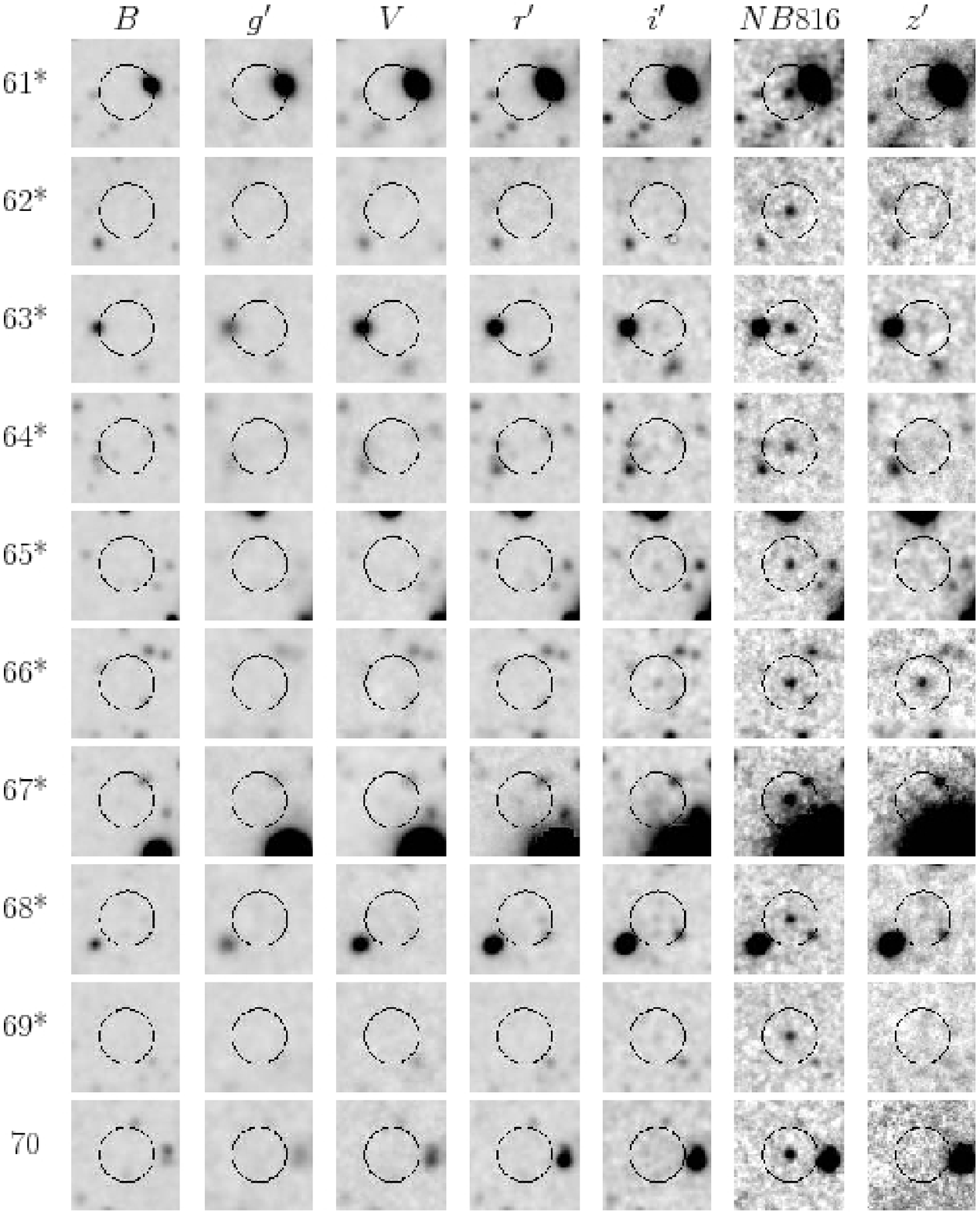}}\\[2mm]
\centerline{Fig. 2. --- continued.}
\clearpage
{\plotone{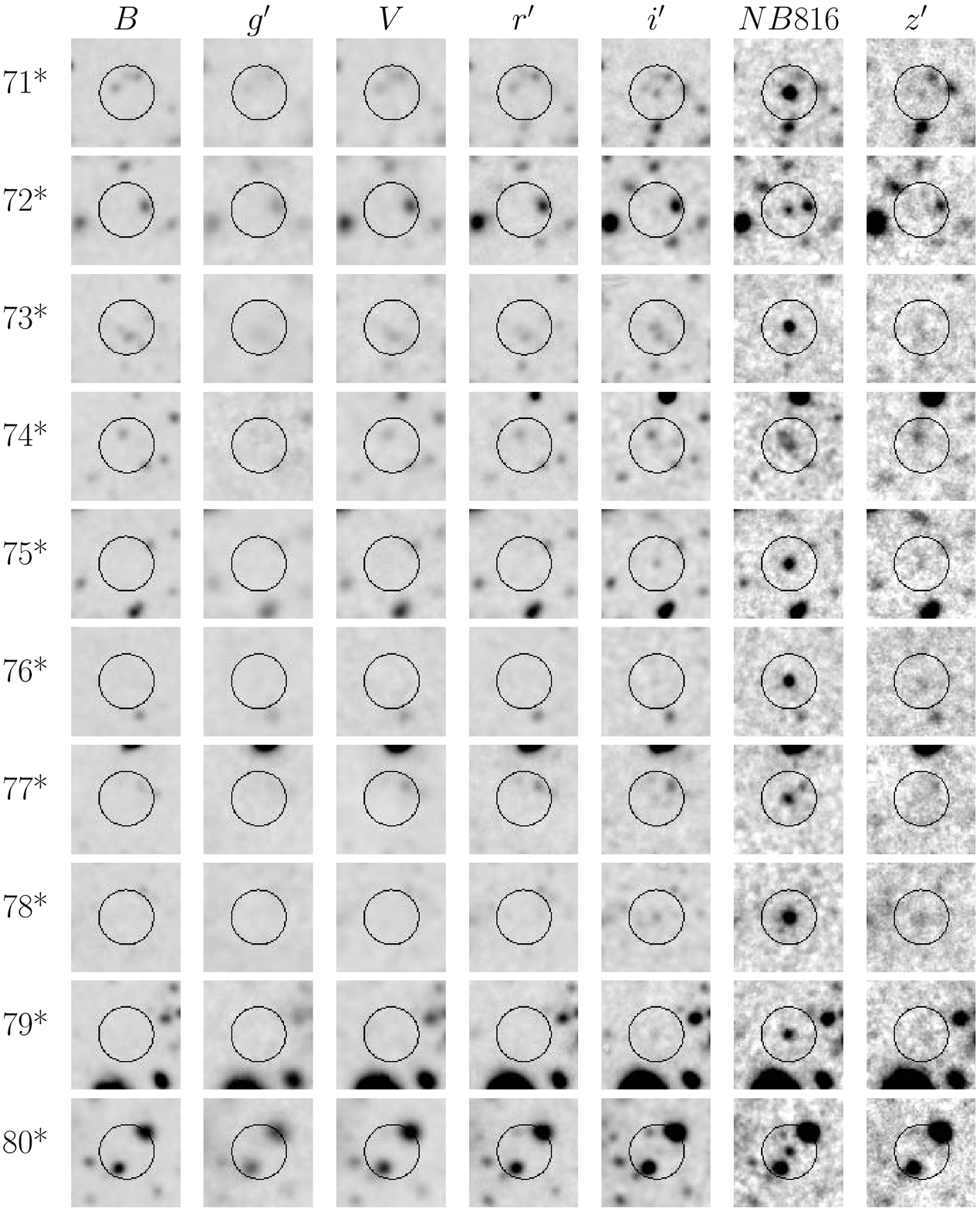}}\\[2mm]
\centerline{Fig. 2. --- continued.}
\clearpage
{\plotone{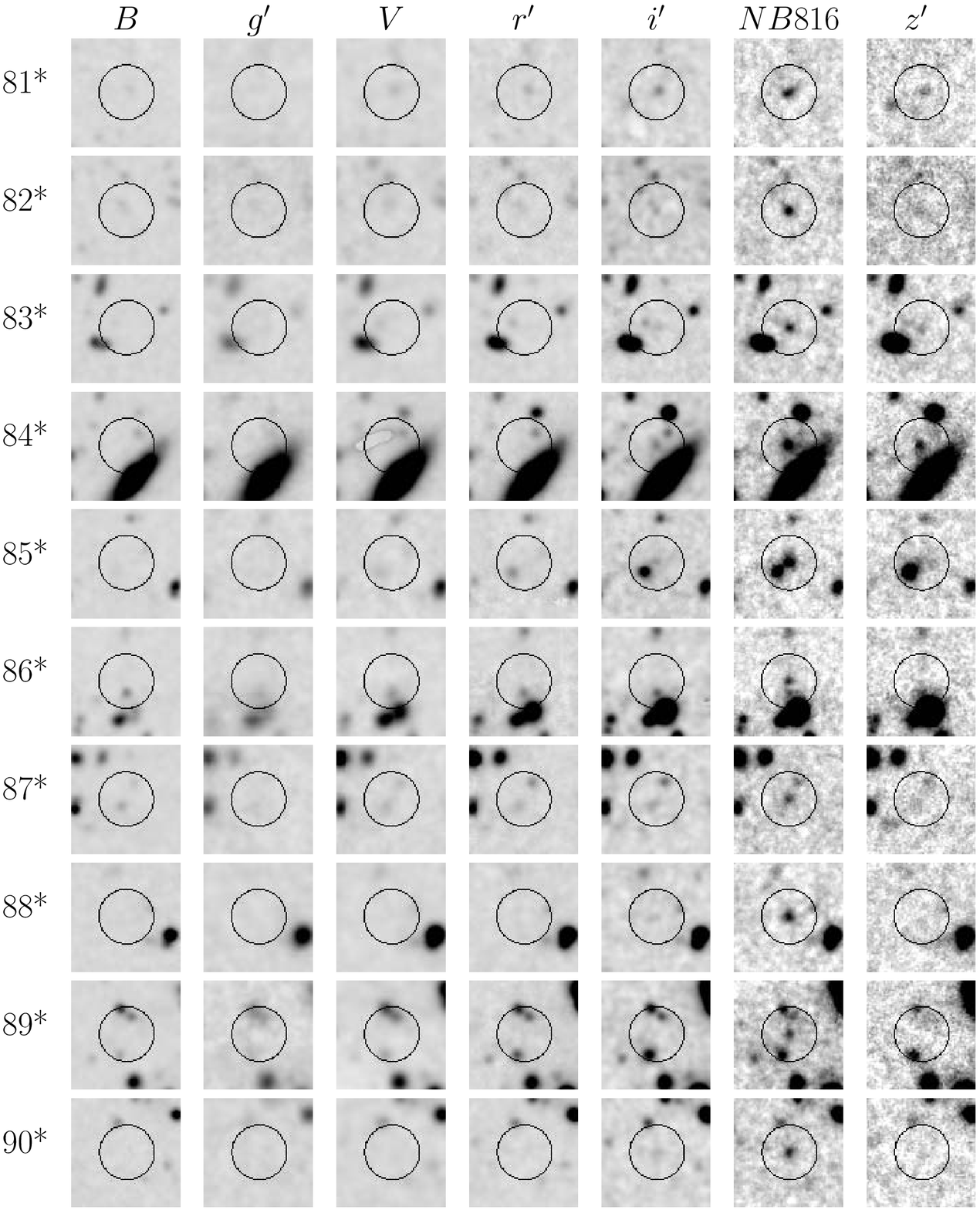}}\\[2mm]
\centerline{Fig. 2. --- continued.}
\clearpage
{\plotone{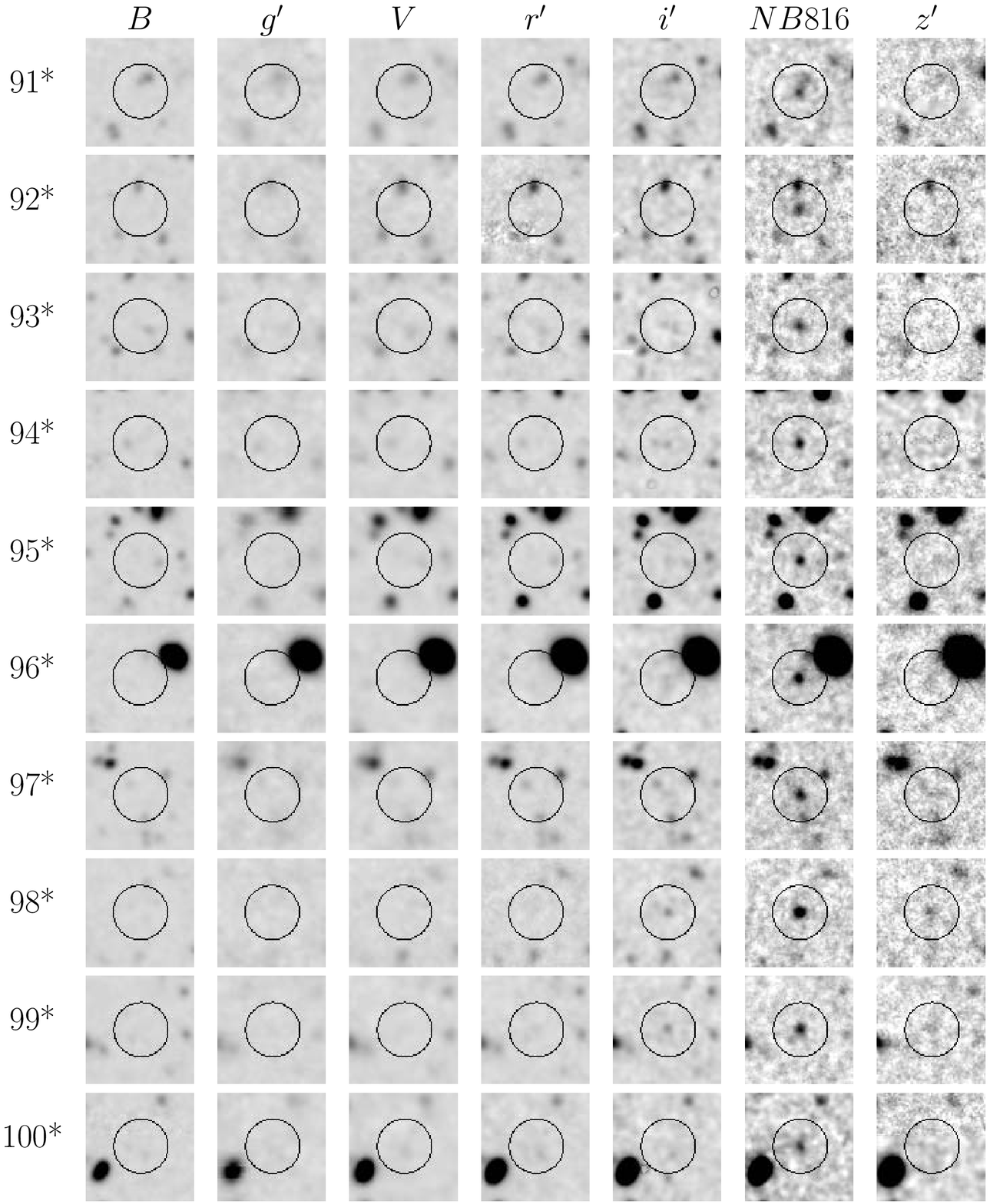}}\\[2mm]
\centerline{Fig. 2. --- continued.}
\clearpage
{\plotone{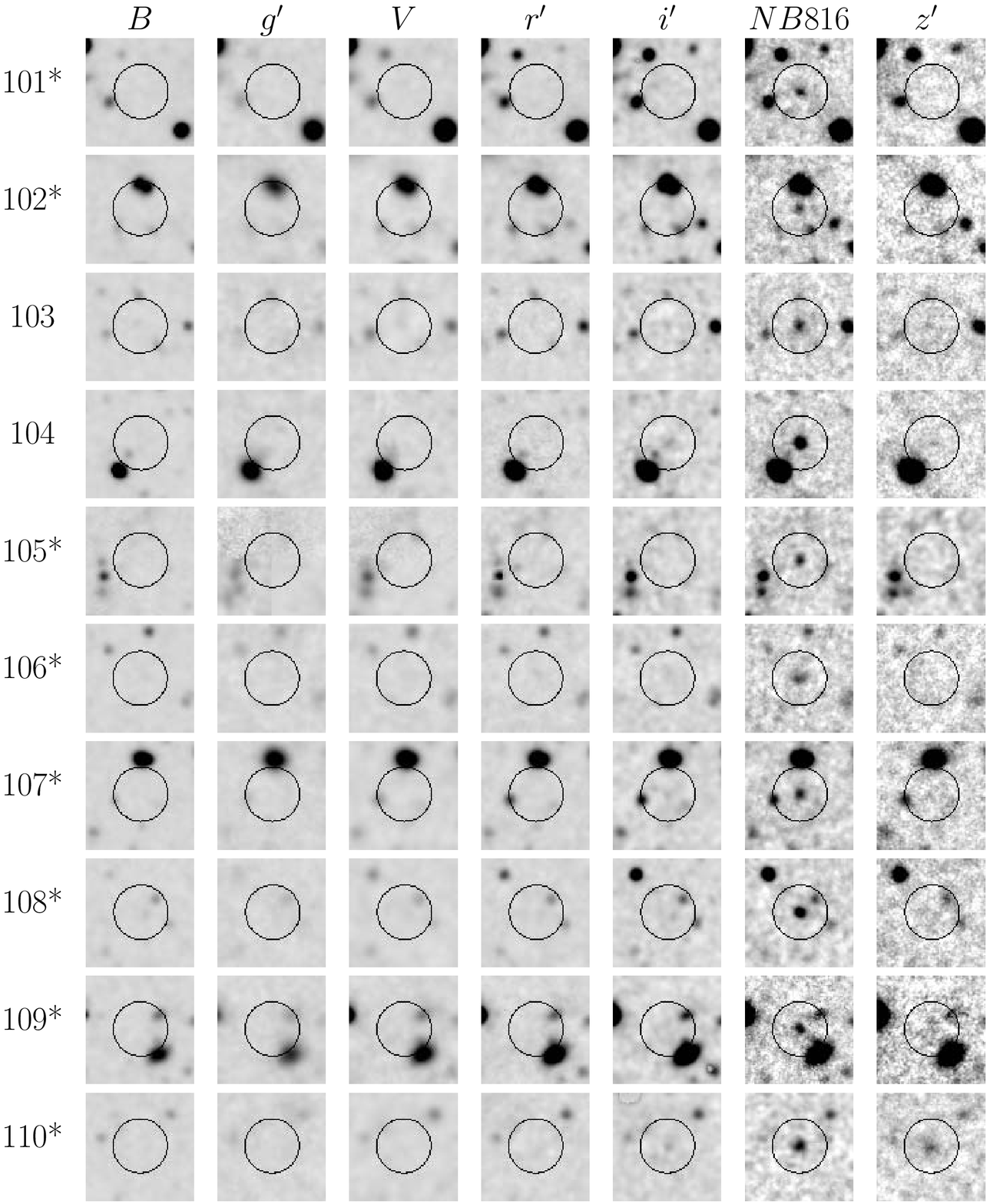}}\\[2mm]
\centerline{Fig. 2. --- continued.}
{\plotone{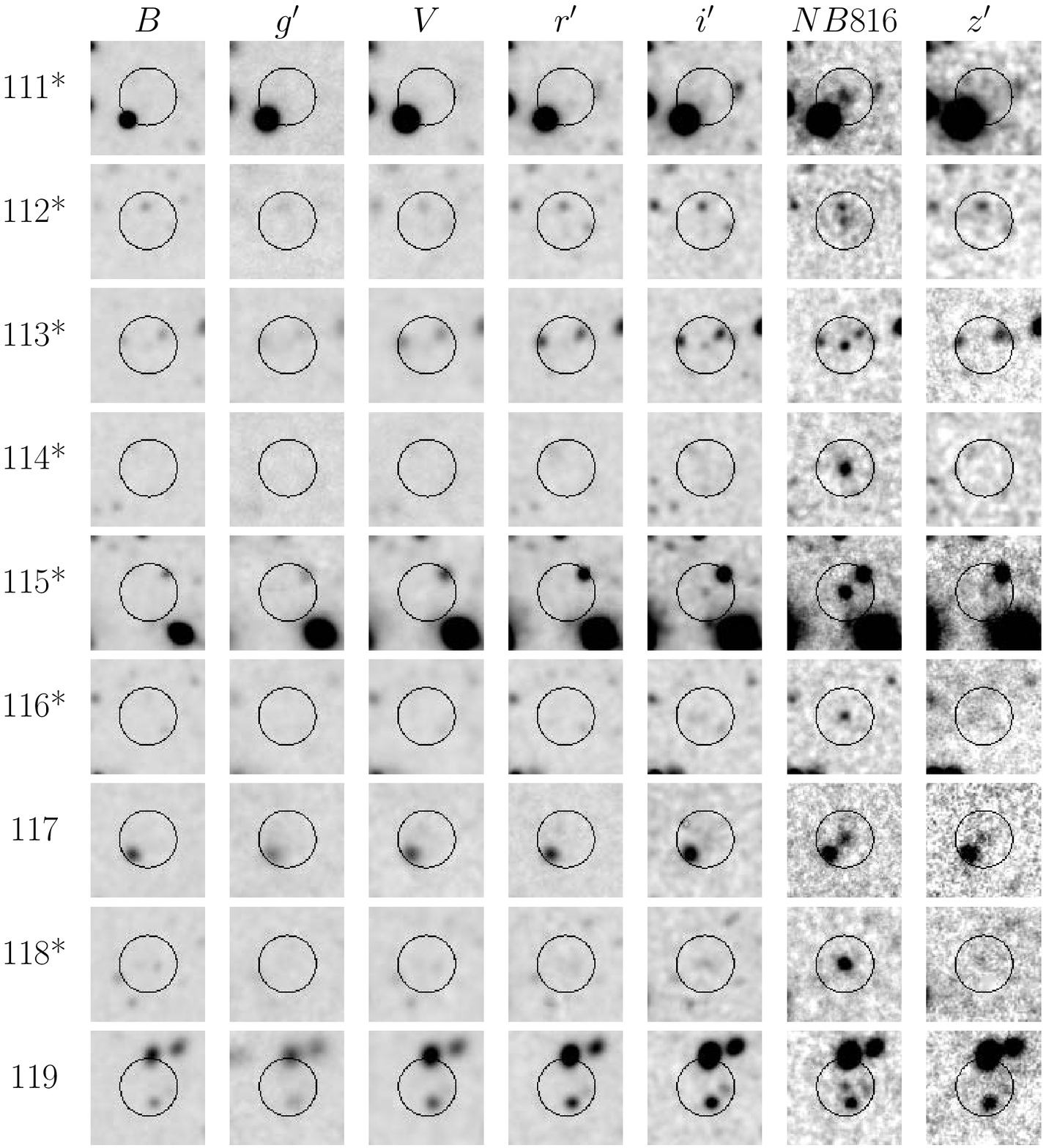}}\\[2mm]
\centerline{Fig. 2. --- continued.}

\clearpage


\begin{figure}
\plotone{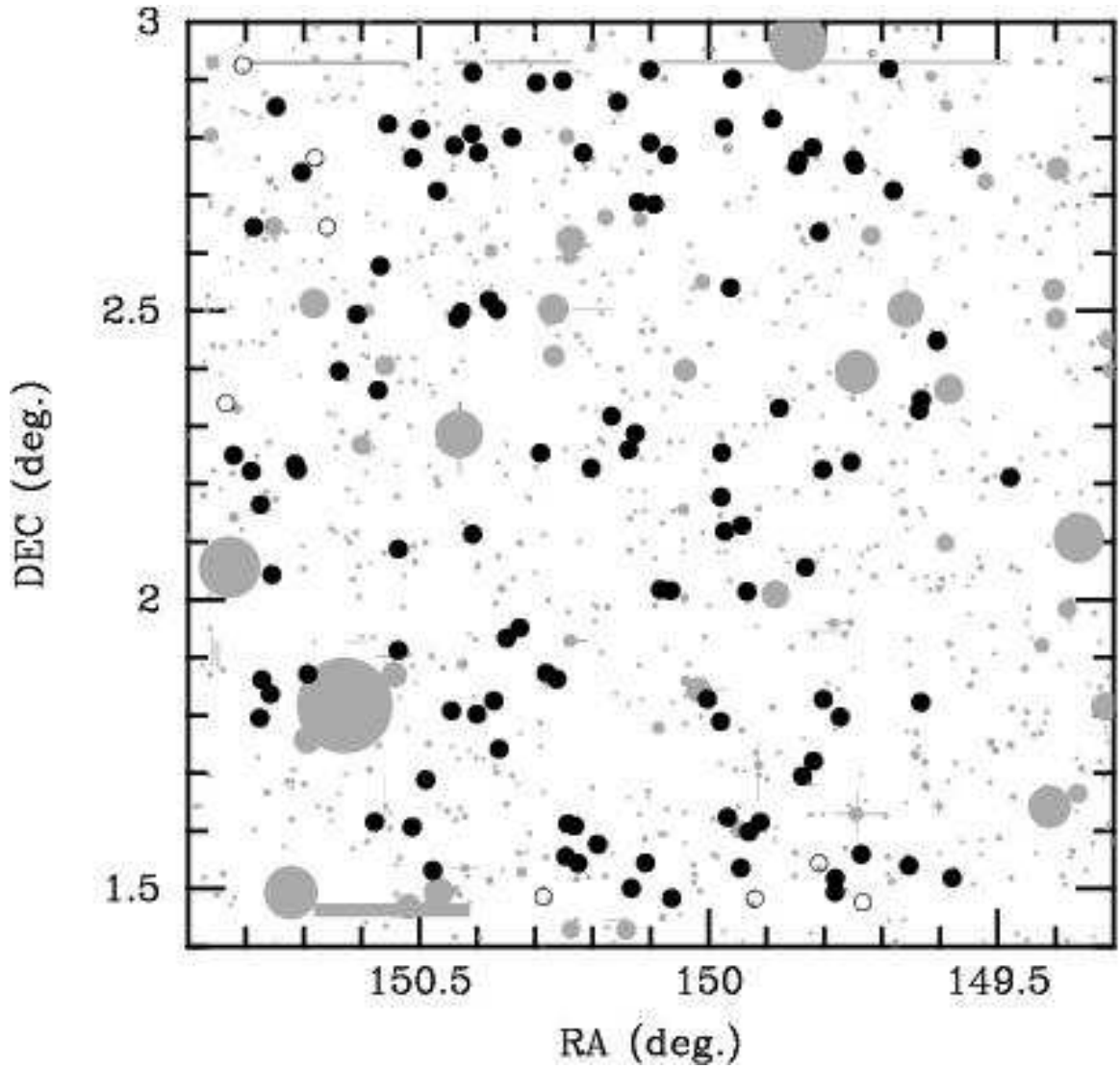}
\caption{Spatial distributions of our 119 LAE candidates.
         The non-statistical sample is shown by open circles.
         The shadowed regions show the areas masked out for the detection.
\label{map}}
\end{figure}
\clearpage

\begin{figure}
\plottwo{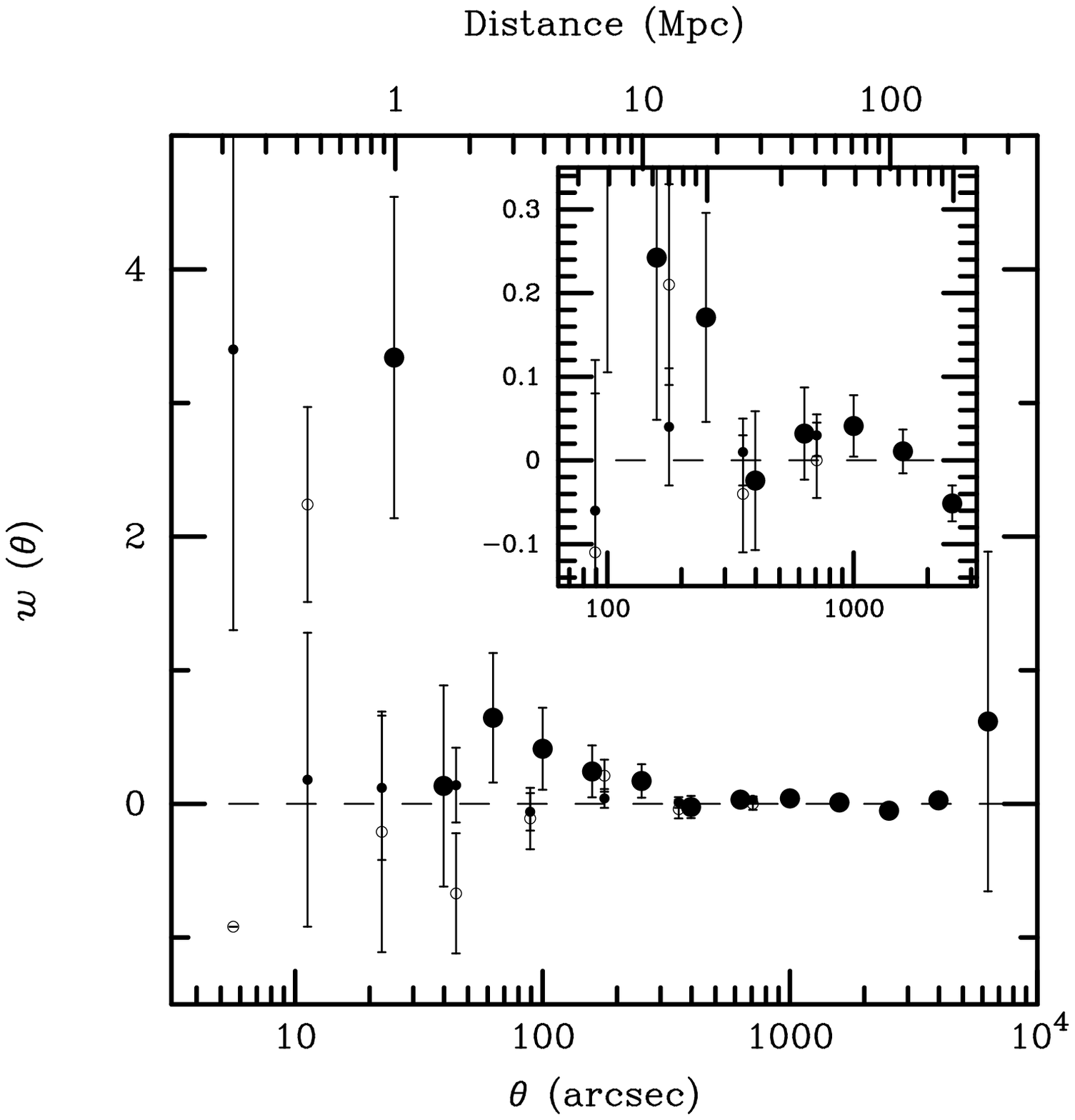}{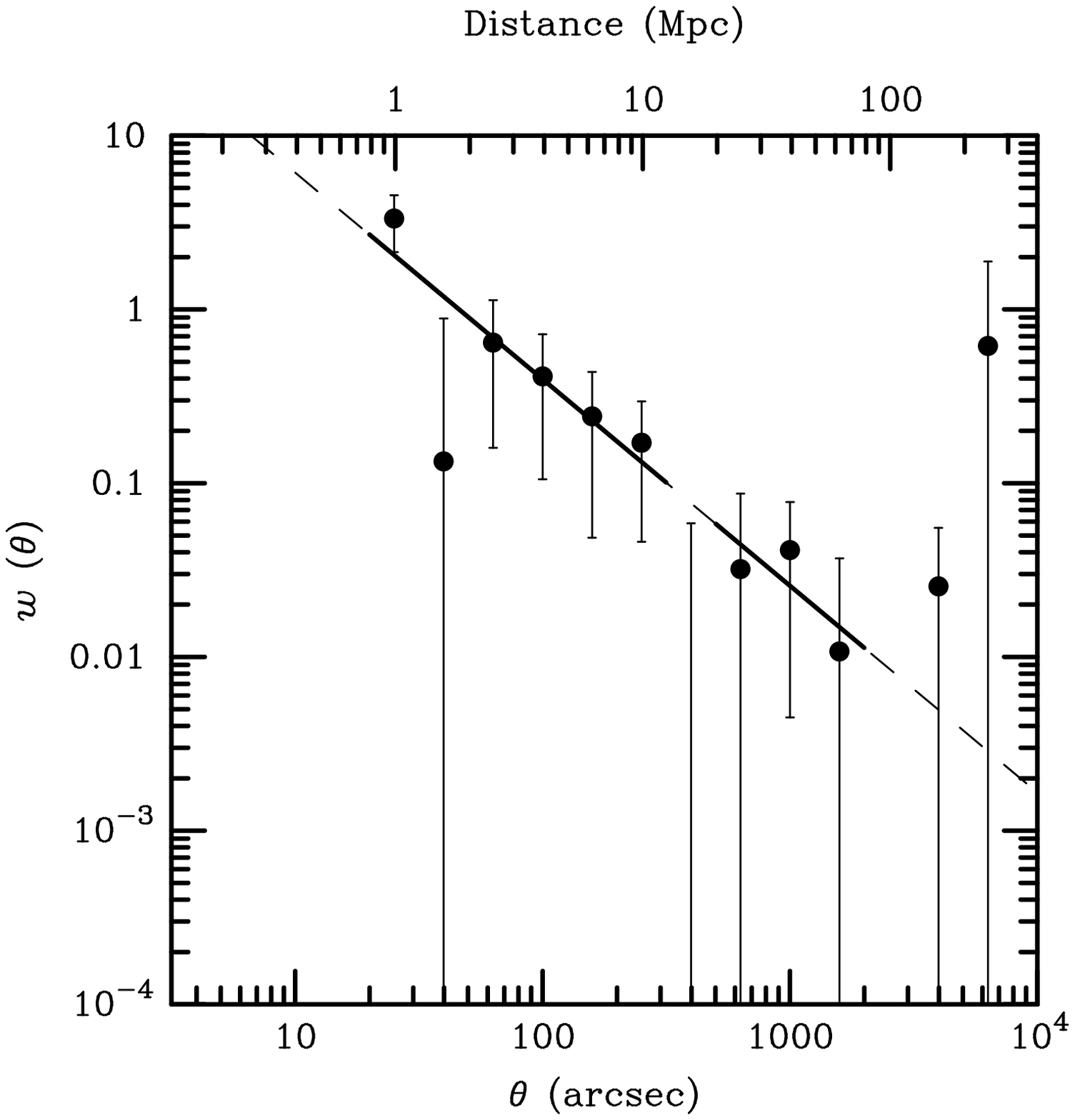}
\caption{
(left panel) Angular two-point correlation function
 of the 111 LAE candidates
in the statistical sample (large filled circles).
The small filled circles correspond to the
 the bright sample (\NB$\le$25.5) of Shimasaku et al. (2006)
while the open circles show 
the  whole sample (\NB$\le$26.5) of Shimasaku et al. (2006).
Note that the \NB{} limit of our LAE candidates is 25.1.
The inset shows an expanded view at 60 arcsec $\le \theta \le 3000$ arcsec.
(right panel) The same as the left panel but $w(\theta)$
is shown by logarithmic scale.
Large filled circles show the angulra two-point correlation function
by the 111 LAE candidates in the statistical sample.
The dashed line shows the result of power-law fit of the data points
at 25 arcsec $\le \theta \le 1585$ arcsec except the point at 398 arcsec
(the fitting region is shown by the solid line).
\label{acf}}
\end{figure}
\clearpage


\begin{figure}
\plottwo{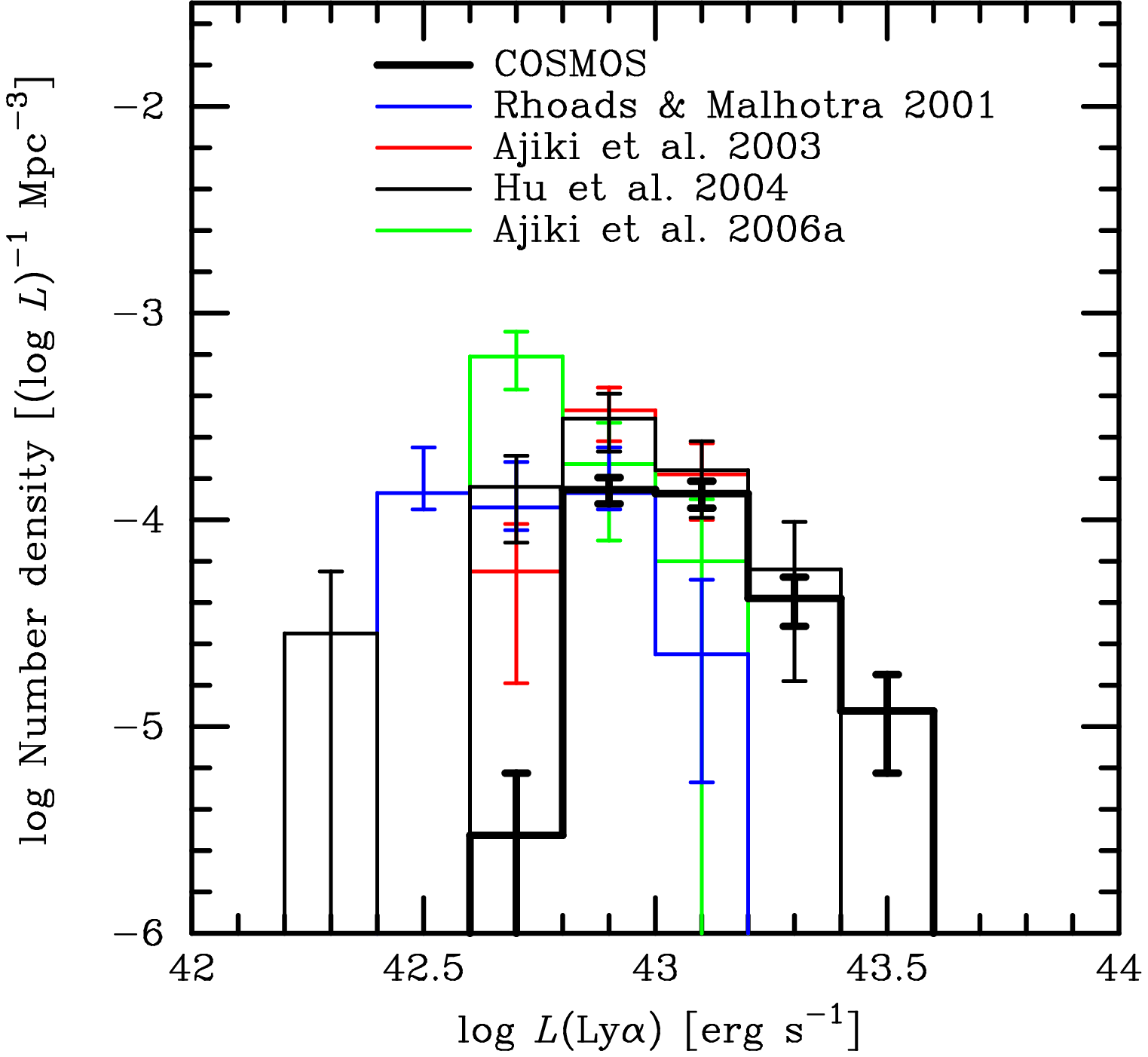}{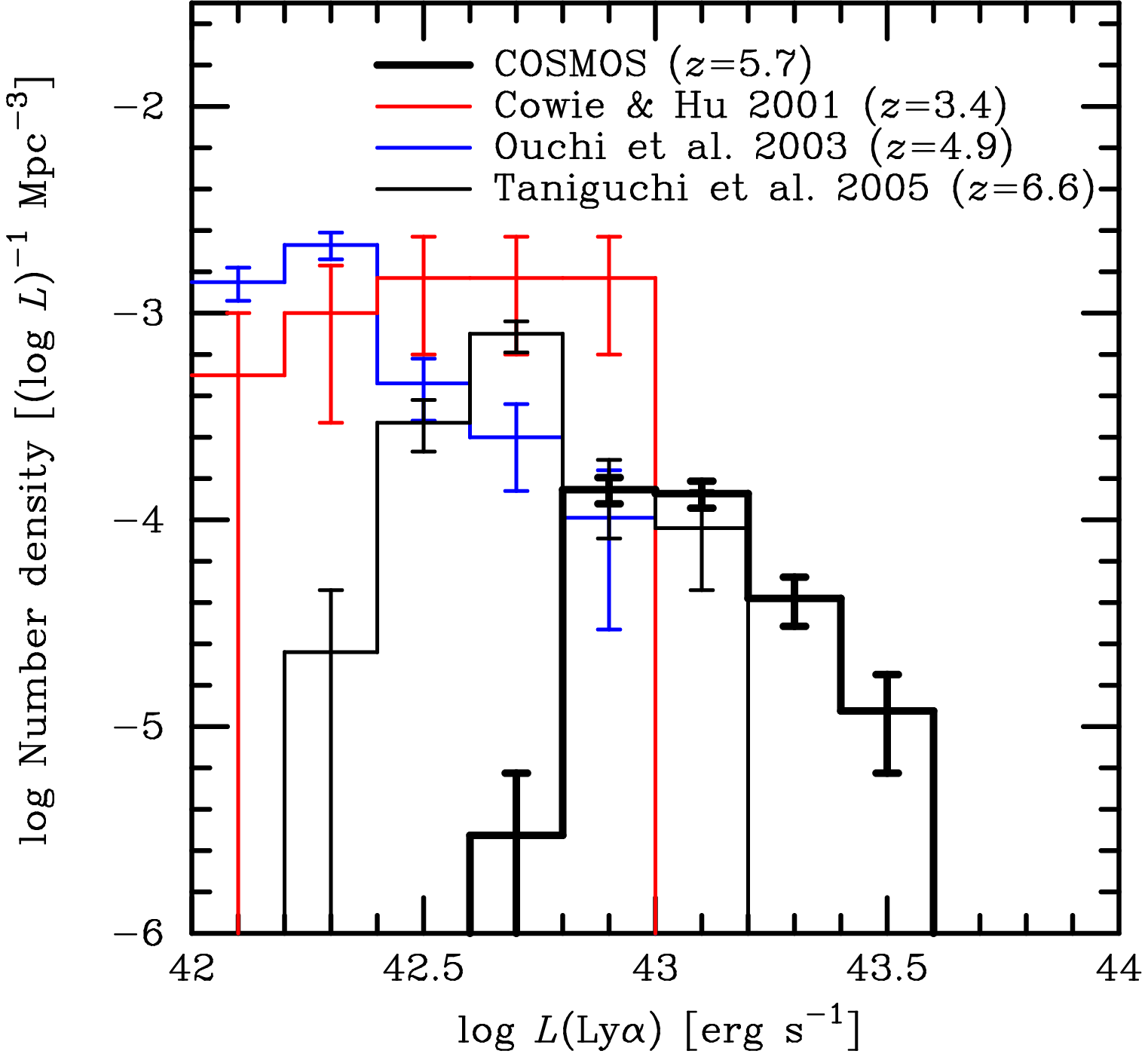}
\caption{Ly$\alpha$ luminosity functions for our LAE candidates.
         (Left panel) Our result (thick line) is compared with those from previous 
         surveys for LAEs  at $z\sim 5.7$  (Rhoads \& Malhotra 2001; Ajiki et al. 
         2003, 2006a; Hu et al. 2004).
         (Right panel) Our result (thick line) is compared with those from previous surveys
         for LAEs  at $z=3.4$ (Cowie \& Hu 1998),
          $z=4.9$ (Ouchi et al. 2003), and  $z=6.6$ (Taniguchi et al. 2005).
\label{ll}}
\end{figure}
\clearpage

\begin{figure}
\plotone{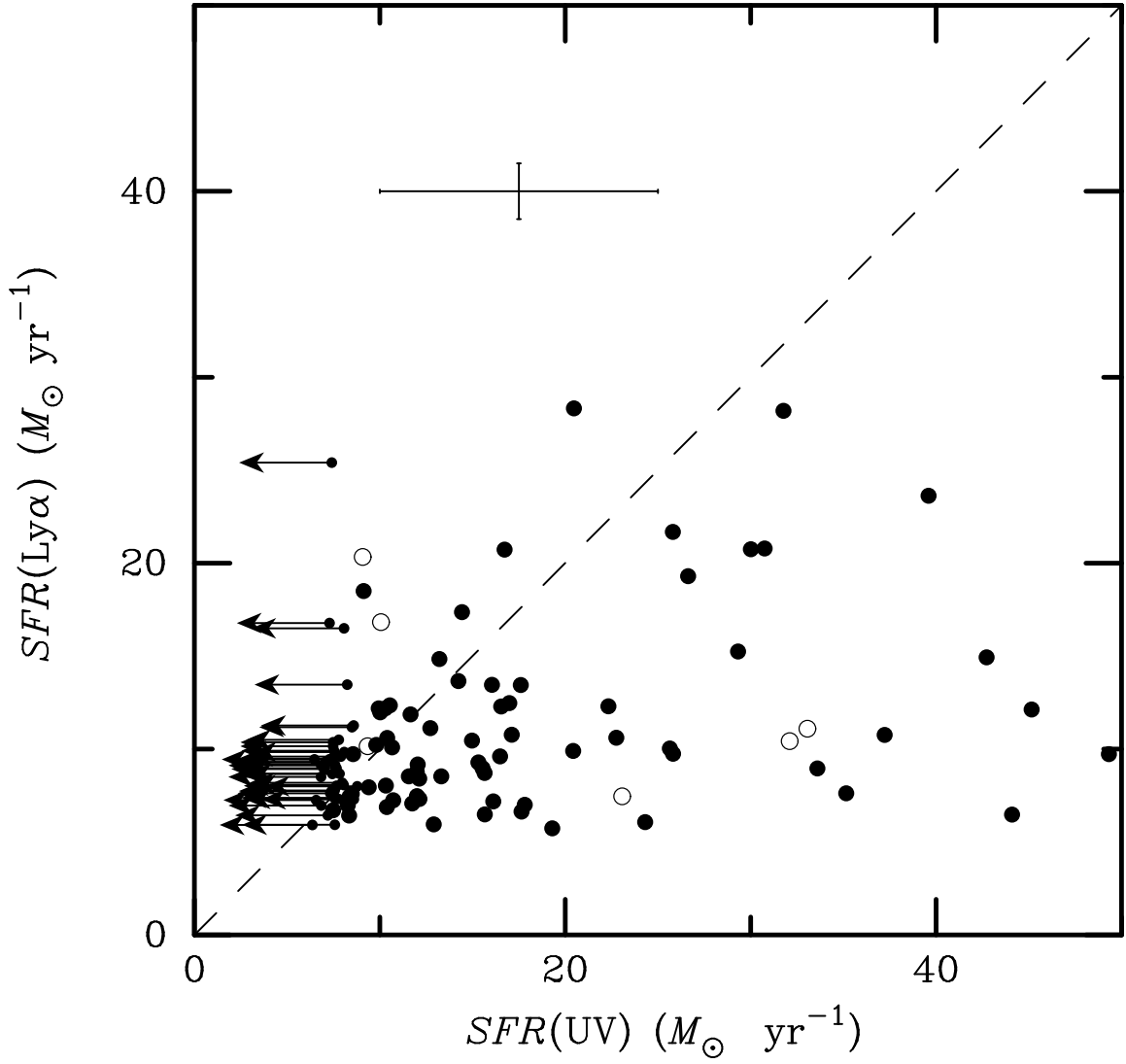}
\caption{Comparison between $SFR$(Ly$\alpha$) and $SFR$(UV) for our 119 LAE candidates. 
         The non-statistical sample is shown by open circles.
         The dashed line shows the relation of $SFR($Ly$\alpha) /SFR($UV$) =1$.
         The cross bar in upper left shows the typical $1 \sigma$ error for our LAE candidates.
\label{sfr}}
\end{figure}

\end{document}